\documentclass{aa}

\usepackage{graphicx}
\usepackage{txfonts}
\usepackage{comment}
\usepackage{hyperref}
\usepackage{amsmath}
\usepackage{enumitem}
\usepackage{gensymb}
\usepackage{xcolor}
\usepackage[normalem]{ulem}
\usepackage{natbib}
\usepackage{placeins}
\usepackage{comment}

\begin{document}

   \title{Particle acceleration signatures in the time-dependent one-zone synchrotron self-Compton model of blazar flares}
   
   \titlerunning{Particle acceleration signatures in the time-dependent one-zone SSC model of blazar flares}

   \author{Paloma Thevenet
           \inst{1}
           \and
          Andreas Zech
          \inst{1}
          \and
          Catherine Boisson
          \inst{1}
          \and
          Anton Dmytriiev
          \inst{2}
          }

   \institute{LUX, Observatoire de Paris, Université PSL, CNRS, Sorbonne Université, 92190 Meudon, France\\
              \email{paloma.thevenet@observatoiredeparis.psl.eu, andreas.zech@observatoiredeparis.psl.eu}
              \and
              Centre for Space Research, North-West University, Potchefstroom 2520, South Africa
             }
 
  \abstract
   {The study of multiwavelength flux and spectral variations during rapid flares from blazars provides strong constraints on the physical parameters of the compact emission regions responsible for these still poorly understood events. Although a full description of the continuous and transient emission from blazars seems to require more sophisticated scenarios, standard leptonic one-zone models are a promising first step toward understanding their underlying physical acceleration and emission processes, when concentrating on the variable emission during rapid flare events.}
   {
    Different scenarios of particle acceleration and loss mechanisms can be approximately described within the simple one-zone framework, enabling a systematic study of their impact on the observable properties of multiwavelength flare light curves. Our goal is to identify characteristic signatures in these light curve profiles that permit one to discriminate between the main physical processes situated inside the relativistic jet and commonly invoked to explain blazar flares. The present study exclusively focuses on modeling flares from BL Lac type objects, which can be described within the synchrotron self-Compton (SSC) emission scenario.
   }
   {Combinations of several commonly employed mechanisms to describe the gain and loss of energetic particles in one-zone models during flaring events are studied in a systematic way: particle injection; diffusive shock and stochastic acceleration and reacceleration; particle escape; adiabatic losses; radiative losses through synchrotron and inverse-Compton radiation. The current study is limited to the case of "hard-sphere" scattering. For each scenario, the resulting multiwavelength light curves are characterized by the shape and timescales of the rising and falling parts of the flare, as well as their asymmetry or lack thereof.
   }
   {A large variety of light curve shapes arises from the different scenarios under study. Characteristic signatures, in particular energy-dependent time delays and differences in the shapes of the rising part of the flare, should allow the distinction to be made between different injection and acceleration scenarios, given the availability of sufficiently high-quality multiwavelength data sets. This is illustrated with a simplified application to a flare event from the blazar Mrk\,421.}
   {}


   \maketitle

\section{Introduction}

Blazars, active galactic nuclei (AGNs) with relativistic jets closely aligned with our line of sight, are characterized by variations in flux, spectrum, and polarization \citep{Urry1995, Padovani1995, Fossati1998}.
  These variations span the entire electromagnetic spectrum and occur over a large range of timescales, indicating a variety of underlying physical processes.
  While flux variations at radio frequencies generally occur on the scale of years or months, flares at the highest observed gamma-ray energies have been detected with much shorter flux doubling times, down to only a few minutes in the most extreme cases \citep[e.g.,][]{Abramowski2010, Albert2007}, pointing to a very compact emission region with large Doppler boosting.
  Even on timescales of hours or days, the physical origin of this rapid variability is not fully understood. Current studies support two main hypotheses: observed flux variability is either primarily attributed to macroscopic changes in the emission region within the jet, or to changes in the energy distribution of the nonthermal particle population responsible for the continuous broadband emission.

The former scenarios include variations in source extension, particle number density, or magnetic field strength, which are related to the geometry and dynamics of the jet \citep[e.g.,][]{Potter2012,Joshi2016,Luashvili2023},  or variations in the pointing direction of the jet, and therefore also in Doppler beaming \citep[e.g.,][]{Casadio2015,Larionov2016,Raiteri2017,Britzen2018}.  
Although these macroscopic changes may not explain the fastest flares, they can still lead to rapid flux variability on the scale of days, depending on the strength of the relativistic beaming involved.

Scenarios that focus on changes of the particle distribution generally invoke different kinds of particle injection, acceleration, and cooling mechanisms to explain rapid flares on the scale of days, hours, or even below. The main acceleration mechanisms in this context are shock acceleration \citep[e.g.,][]{Marscher1985,Sikora2001,Boettcher2019, Lemoine2019a, Lemoine2019b, Lemoine2019c}, acceleration on turbulences \citep[e.g.,][]{Boutelier2008,Tammi2009,Tramacere2011, Lemoine2020}, or magnetic reconnection \citep[e.g.,][]{Giannios2009,Gouveia2010,Shukla2020,Geng2022}.

Current studies of multiwavelength (MWL) variability typically focus either on the long-term stochastic or, in some candidate sources, the quasi-periodic behavior of light curves \citep[LC, e.g.,][and references therein]{Bhatta2020,Goyal2022,OteroSantos2024,deJaeger2023,Covino2019}, or on the investigation of specific, significant events characterized by rapid flux variations, such as bright blazar flares \citep[e.g.,][]{Palyia2015,Prince2018,EscuderoPedrosa2024}. These two approaches are complementary in probing the underlying physical mechanisms, either through statistical analyses of variability patterns, such as the precise shape of the power spectral density, or through detailed radiative modeling of individual flaring episodes. This two-pronged approach seems even more justified when considering that a variety of mechanisms provoking flux variations may be at play even in a single source \citep[e.g.,][]{Matthews2020}.

Individual source studies, for example of events including the 2010 Mrk\,421 flare \citep{Abeysekara2020}, multiple flares of PKS\,2155-304 in 2006 \citep{Abramowski2012}, and flares of 3C\,279 observed with H.E.S.S. \citep{Abdalla2019}, suggest that the simplest emission model, the one-zone synchrotron self-Compton (SSC) model, has difficulty explaining the multiwavelength flux variations from a more or less steady baseline flux to the flare state. In general, it may be necessary to invoke one or more distinct emission regions to account for the full blazar emission, including both the continuous emission with stochastic fluctuations and the aperiodic flare events. Impressive progress is being made on sophisticated multi-zone models to describe the observed stochastic patterns of flux variability \citep[e.g.,][and references therein]{Roeken2018,DiGesu2022,Jormanainen2023,SolZech2022}. 

However, when only focusing on the flux variations above the baseline during the flare itself, a radiative model based on a single homogeneous emission region should provide a reasonable approximation for rapid flares, given the compactness of the emission inferred from the variability timescale. This approach has the advantage of permitting us to focus on the physical mechanisms expected to be at play at these short timescales, while ignoring the macroscopic details of the emission region.

Blazars indeed display a striking diversity of flaring behavior across the electromagnetic spectrum. The LCs can differ markedly in duration, symmetry, and spectral evolution, including fast-rise and slow-decay, slow-rise and fast-decay, and nearly symmetric profiles  \citep[e.g.,][]{Nalewajko2013,Kushwaha2014, Kapanadze2025, Khatoon2024, Cerruti2025}, sometimes featuring an apparent plateau instead of a sharp peak \citep[e.g.,][]{Rajput2019,Yu2024}. When considering different model approaches, it should be kept in mind that the number of well-sampled MWL flares on short hour-to-day timescales remains limited. Detailed studies of correlated flux variability in blazars are often hindered by the poor and mostly irregular sampling of the observed LCs, as well as integration times that vary significantly between instruments. This motivates a study of simplified scenarios, where the focus is not on reproducing a single event in detail but on identifying generic observational signatures of different physical processes.

In this paper, we focus thus on a purely leptonic time-dependent SSC model that does not include external photon fields, a framework well adapted to most BL Lac type blazars. We consider events of short duration, constraining the emission region to a compact zone due to causality arguments. Our aim is to identify observable signatures in the shape of the resulting LCs and in the relative amplitudes between different bands that could help distinguish between the generic flaring scenarios under study. While the physical characteristics of steady-state solutions in such a model have been described for a range of injection and acceleration scenarios \citep[e.g.,][]{Kirk1998,Katarzynski2006,Tramacere2011,Baring2017,Lewis2018} to match the observed spectral energy distributions, a systematic study of multiwavelength flux variability for different mechanisms is still needed.
As steady-state solutions for the emission from BL Lac objects usually indicate environments of low magnetization, we limit the present study to acceleration mechanisms of the Fermi type and do not address magnetic reconnection scenarios, which are evoked in particular to explain very rapid minute-scale flares \citep[e.g.,][]{Giannios2009, Nalewajko2011}, arising probably from regions of high magnetization close to the base of the jet.

The paper is organized as follows. A description of the emission model and the different scenarios that are investigated is given in Section~\ref{sec::model}. The systematic exploration of the different physical scenarios and their impact on the resulting SED and LCs is described in Section~\ref{sec::lightcurves}. A comparison of the obtained LCs under different assumptions allows us to extract observable features to distinguish between the different scenarios, given a sufficiently precise multiwavelength coverage (Section~\ref{sec::discussion}). To illustrate the differences in the predicted flux evolution for the most efficient flaring scenarios in a concrete case study, we compare them with an observed flare from Mrk\,421 in 2013 in Section~\ref{sec::application}. We summarize our findings and give an outlook in Section \ref{sec::conclusions}.

\section{Emission model setup}
\label{sec::model}

To simulate rapid flares, we use the EMBLEM (Evolutionary Modeling of BLob EMission) code, developed by \cite{Dmytriiev2021}.
We assume a blob-in-jet model and evolve the electron population of a single plasma blob with a tangled magnetic field $B(t)$. We denote its radius $R(t)$, Doppler factor $\delta$ and redshift $z$. The blob continuously undergoes particle injection $\Dot{Q}_{\rm inj}(\gamma,t)$, balanced by radiative cooling and escape to reach a steady state. During the flaring phase, a time-dependent injection can cause a flux variation on top of the quiescent steady-state emission. 

Additionally, the electrons in the blob can be accelerated or reaccelerated by two mechanisms: stochastic acceleration (Fermi II) and diffusive shock acceleration (DSA, Fermi I) which are parameterized by characteristic acceleration timescales. 
The particles lose energy by emitting synchrotron and inverse-Compton (IC) radiation in the framework of the SSC scenario and they escape the blob within a timescale $t_{\rm esc}$, which is independent of time and energy unless turbulence is present. In the latter case, we denote the escape timescale $t_{\rm esc}^{\rm (turb)}(\gamma,t)$. 
The initial EMBLEM code has been improved by including a treatment of adiabatic expansion, with a timescale $t_{\rm ad}$, and energy- and time-dependent timescales for the case of stochastic acceleration.

The evolution of the electron distribution $N_e(\gamma,t)$ in the blob is governed by a Fokker-Planck equation \citep[e.g.,][]{Kardashev1962, Tramacere2011}: 
\begin{align}
    \frac{\partial N_e(\gamma,t)}{\partial t} = \frac{\partial}{\partial \gamma}\left[\left(b_c(\gamma,t)\gamma^2+\frac{1}{t_{\rm ad}}\gamma-a(t)\gamma-\frac{2}{\gamma} D_{\rm F_{II}}(\gamma,t)\right)N_e(\gamma,t)\right] \nonumber\\
    +\frac{\partial}{\partial \gamma}\left(D_{\rm F_{II}}(\gamma,t)\frac{\partial N_e(\gamma,t)}{\partial \gamma}\right) 
    -N_e(\gamma,t)\left(\frac{1}{t_{\rm esc}}+\frac{3}{t_{\rm ad}}\right)+\Dot{Q}_{\rm inj}(\gamma,t).  \label{Fok_P_eq1}   
\end{align}

The code solves the kinetic equation using a fully implicit difference scheme detailed in \cite{Chang1970} and described in Appendix \ref{app:numerical}.

\subsection{Radiative and adiabatic cooling}
The radiative cooling rate $b_c(\gamma,t)$ takes into account both the synchrotron and IC cooling, $-b_c\gamma^2=\Dot{\gamma}_{S}+\Dot{\gamma}_{IC}$ where $\Dot{\gamma}$ denotes the time derivative of the Lorentz factor. The details of the synchrotron and IC cooling implementation can be found in \cite{Dmytriiev2021}.
\begin{comment}
The rate of synchrotron cooling, caused by the synchrotron radiation of high-energy electrons, is given by:
\begin{equation}
-\Dot{\gamma}_{S}=\frac{4\sigma_T}{3m_ec}\gamma^2 U_B    
\end{equation}
with $\sigma_T$ the Thomson cross-section, $m_e$ the electron mass and $U_B=B^2/2\mu_0$ the magnetic energy density, where $\mu_0$ denotes the magnetic permeability. The rate of IC cooling, resulting from the upscattering of synchrotron photons by high-energy electrons, can be expressed as \citep{Moderski2005}:
\begin{equation}
-\Dot{\gamma}_{IC}=\frac{4\sigma_T}{3m_ec}\gamma^2\int_{\epsilon_{\rm min}}^{{\epsilon_{\rm max}}}f_{\rm KN}(4\gamma\epsilon)u_Sd\epsilon
\end{equation}
where $u_S$ is the energy distribution of the ambient photons and $f_{\rm KN}(x)$ includes the Klein-Nishina IC cross-section.
\end{comment}
We treat the IC cooling as a continuous-loss process, neglecting the discrete losses, which appear to be significant only for extreme gamma-ray flares of FSRQ objects \citep[e.g.,][]{Dmytriiev2024}.
        
Adiabatic expansion affects the electron energy distribution in two manners \citep[e.g.,][]{Gould1975}. First, the relativistic electrons lose energy at a rate $1/t_{\rm ad}$ as they behave kinematically similar to photons. Second, the change in volume of the blob leads to a direct change of the electron spectrum $N_e(\gamma,t)$ because of the decrease in particle density during the expansion, expressed by the term $-3N_e(\gamma,t)/t_{\rm ad}$. The adiabatic cooling timescale is $t_{\rm ad}(t)=R(t)/(\beta_{\rm exp}c)$, where the expansion speed $\beta_{\rm exp}$ depends on the jet velocity $\beta_{\rm jet}$ and  the intrinsic jet opening angle $\alpha$: $\beta_{\rm exp} = \beta_{\rm jet}\tan(\alpha)$.
A study by \citet{Pushkarev2009} on radio very-long-baseline interferometry (VLBI) observations found $\alpha = \rho/\Gamma$ where $\rho \approx 0.26$\, rad. We fix this value in our simulations, while the bulk Lorentz factor and the jet velocity are obtained by fixing the viewing angle $\theta$ for a given Doppler factor.
The presence of adiabatic expansion implies an evolving blob radius and magnetic field \citep{Tramacere2022}:
\begin{equation}
    R(t) = R_0 + \beta_{\rm exp}c(t-t_{\rm exp})H(t-t_{\rm exp}),
\end{equation}
\vspace{-0.7 cm}
\begin{equation}
    B(t)=B_0\left(\frac{R_0}{R(t)}\right)^{m_B},
\end{equation}
where we assume that the expansion of the blob begins at $t=t_{\rm exp}$ and $H$ is the Heaviside function, with $R_0$ and $B_0$ the initial radius and magnetic field, and $m_B$ an index indicating the geometric configuration of the magnetic field: $m_B=1$ for the toroidal configuration that we consider in this study \citep[e.g.,][]{Begelman1984}.

\subsection{Particle injection and (re)acceleration}

The last term on the right-hand side (rhs) of Equ.~\ref{Fok_P_eq1} accounts for particle injection (in units of cm$^{-3}$s$^{-1}$). The electron distribution for the quiescent and/or flaring injection follows a power law with an exponential cutoff:
\begin{equation}
    \Dot{Q}_{\rm inj}(\gamma)=\Dot{N}_{\rm inj}\left(\frac{\gamma}{\gamma_{\rm inj,pivot}}\right)^{\alpha_{\rm inj}}\exp\left(-\frac{\gamma}{\gamma_{\rm inj,cut}}\right),
\end{equation}

where $\Dot{N}_{\rm inj}$ is the spectrum normalization, $\gamma_{\rm inj,pivot}$ the pivot Lorentz factor, $\alpha_{\rm inj}$ the spectrum slope and $\gamma_{\rm inj,cut}$ the cutoff Lorentz factor.

In addition to the simple injection scenario, two particle acceleration and two reacceleration scenarios 
have been implemented, counting among the most basic and widely applied scenarios to explain variability in blazars. 
The electrons are accelerated during the flaring phase through the Fermi I or Fermi II mechanism, starting from a thermal or mildly relativistic distribution. By reacceleration we understand that the continuously injected electrons in the blob, which we assume have been accelerated to a power law distribution before entering the emission region, are being additionally accelerated inside the blob through the Fermi I or Fermi II mechanism during the flaring phase. Such reacceleration mechanisms are relevant to explain the very hard $\gamma$-ray spectra from "extreme blazars," also known as ultra-high frequency peaked BL Lac objects \citep{Biteau2020, Tavecchio2022, Zech2021}.

In the case of Fermi I acceleration on a shock in a purely leptonic plasma, a power-law tail of accelerated particles emerges from a population of thermal or mildly relativistic particles.
A power-law with exponential cutoff and with a time dependent maximum Lorentz factor is assumed to mimic a shock front injecting particles accelerated inside the blob:
\begin{equation}
    \Dot{Q}_{\rm add, FI}(\gamma)=\Dot{N}_{\rm add}\left(\frac{\gamma}{\gamma_{\rm add,pivot}}\right)^{\alpha_{\rm add}}\exp\left(-\frac{\gamma}{\gamma_{\rm add,cut}(t)}\right).
\end{equation}
Unlike in the simple particle injection flare scenario, the cutoff Lorentz factor is no longer a fixed value
and is expressed as \citep{Kirk1998} :
\begin{equation}
    \gamma_{\rm add,cut}(t) = \left[\frac{1}{\gamma_{\rm max}}+\left(\frac{1}{\gamma_{\rm add,min}}-\frac{1}{\gamma_{\rm max}}\right)e^{-t/t_{\rm shock}}\right]^{-1},
\end{equation}
with the maximum Lorentz factor:
\begin{equation}
    \gamma_{\rm max} = (\beta_s t_{\rm shock})^{-1},\\
    \beta_s = \frac{4\sigma_T}{3m_e c}\left(\frac{B(t)^2}{2\mu_0}\right).
\end{equation}

While \citet{Kirk1998} and \citet{Kusunose2000} model an acceleration region separate from the emission region and with a negligible contribution to the radiative emission, we inject directly the distribution of accelerated electrons, following their solution, into our emission region, without explicitly modeling a separate acceleration region. Following \citet{Kusunose2000}, in our generic one-zone approach, we do not consider a specific shock geometry and neglect the gradual filling of the emission region with accelerated particles that would be expected for example from a plane shock front moving through the region.

We note that the evolution of the maximum Lorentz factor depends on the shock acceleration timescale as well as on the magnetic field of the environment, which decreases in case of an adiabatically expanding blob.
Different from the above applications, we treat this acceleration timescale as a free parameter that can take on values larger than $R/c$.

To keep the approach deliberately simple, we limit the description to an energy-independent timescale, assuming the limit of hard-sphere scattering
\citep[e.g.,][]{Kirk1998, Kusunose2000, Asano2018} for all of the acceleration and escape processes explored in this first study. We discuss some energy-dependent extensions in section \ref{sec::discussion}.
In the same sense, spatial dependence is neglected and our approach does not capture various DSA-specific processes such as self-generated turbulence or the backreaction of particle pressure on the shock structure, the interest of the present study being to predict whether already the most basic scenarios might be distinguishable with currently available flare data sets.

The relativistic particles in the blob can also be reaccelerated through consecutive shocks. Successive shock encounters can, under certain conditions, harden the spectrum and increase the minimum Lorentz factor of the electron population \citep[][and references therein]{Zech2021}. 
Our treatment of Fermi-I reacceleration in Equ.~\ref{Fok_P_eq1} is again very generic and simplified. 
To mimic the expected spectral evolution from shock reacceleration, we use a description that leads to a systematic energy gain of the particles through the term $a(t)=1/t_{\rm F_{I}}$ in Equ.~\ref{Fok_P_eq1} where $t_{\rm F_{I}}$ is the time and energy-independent reacceleration timescale. This is the same approach used to describe the "acceleration region" by \citet{Kirk1998}, only that here the acceleration term is not applied to particles injected at low energies, but to an already relativistic particle population.
The numerical treatment requires nonzero values for the terms linked to stochastic acceleration (see Equ.~\ref{equ:t_acc_fermi2}), representing the presence of some turbulence in the emission region but with a negligible impact on particle acceleration compared to the systematic energy gain. 

Instead of injecting an additional component of accelerated particles as in the initial Fermi I scenario, all the electrons responsible for the steady-state emission are being reaccelerated. The "acceleration region" is thus identical to the "emission region" in this approach.
While this simplified prescription does not capture the full range of complex effects present in more detailed kinetic models, it does qualitatively reproduce the key features expected in mildly relativistic shocks: hardening of the spectrum and increase in the low-energy cutoff (cf. Appendix~\ref{app:multishock}). 
In general, the outcome of reacceleration depends on the slope of the preexisting distribution $s$ and the shock compression ratio 
$r$ and nonlinear effects can limit the degree of spectral hardening that is achievable \citep[e.g.,][]{Melrose1993, Vieu2022}.

The Fermi II process considered here is the stochastic acceleration of electrons through resonant scattering on MHD waves \citep{Tramacere2009, Tramacere2011, Dmytriiev2021}. The fourth rhs term corresponds to the drift of electrons to higher Lorentz factor due to stochastic acceleration, with the diffusion coefficient $D_{\rm F_{II}}(\gamma,t)=p^2/t_{\rm F_{II}}$ where $t_{\rm F_{II}}$ is the Fermi II acceleration timescale and $p$ is the relativistic electron momentum. 
The acceleration timescale is:
\begin{equation}
    t_{\rm F_{II}}= \frac{1}{\beta_A^2} \left(\frac{\delta B}{B(t)}\right)^{-2} \frac{\lambda_{\rm max}}{c}\left(\frac{r_L}{\lambda_{\rm max}}\right)^{2-q}
    \label{equ:t_acc_fermi2},
\end{equation}
with the turbulence level $\mathcal{L}_{\rm turb}=\delta B/B(t)$, the longest wavelength in the Alfvén spectrum $\lambda_{\rm max}$, the slope of the turbulent spectrum $q$, and the electron Larmor radius $r_L$ and Alfvén speed $\beta_A$ defined as in \citet{Dmytriiev2021}.

In the case of hard-sphere scattering ($q=2$) we focus on here, the acceleration timescale becomes independent of the energy.
Stochastic acceleration also leads to the diffusion of the electron distribution with the Lorentz factor which is described by the fifth term of Equ.~\ref{Fok_P_eq1}, right-hand side.
To model initial Fermi II acceleration and reacceleration, we apply the same Fokker-Planck equation either to a thermal population of electrons or to the relativistic electron population.

\subsection{Particle escape}

Particle escape from the emission region is taken into account in all scenarios. The escape timescale $t_{\rm esc}$ is assumed constant for flares resulting from particle injection and from Fermi I acceleration. However, if the blob expands adiabatically, the timescale evolves so that it remains proportional to $R(t)/c$. When turbulence is present, the escape timescale depends on time and energy \citep[e.g.,][]{Tramacere2011}:
\begin{equation}
    t_{\rm esc}^{\rm (turb)}= \left(\frac{R(t)}{c}\right)^2 \left(\frac{\delta B}{B(t)}\right)^2 \frac{c}{\lambda_{\rm max}}\left(\frac{r_L}{\lambda_{\rm max}}\right)^{q-2}.
    \label{equ:tesc_fermi2}
\end{equation}

This expression is again independent of energy in the hard-sphere case we focus on.

\subsection{Light-crossing effect}

We have added a treatment of the light crossing effect in post-processing, with a method largely inspired by the approach in the open-source code {\it Jet Set}\footnote{\url{https://github.com/andreatramacere/jetset}} \citep[e.g.,][]{Tramacere2009, Tramacere2011, Tramacere2020}. To account for the time delays between photons emitted at different points in the transparent spherical emission region, the latter was sliced orthogonally to the direction of the observer into spherical segments with a thickness corresponding to the light-travel distance for one numerical time step. At each time step, the observed light flux is thus the sum of light emitted from different slices at different times. The contribution of a slice to the flux is normalized by its 
relative volumetric contribution to the sphere. The treatment also accounts for the
changing volume of the blob during adiabatic expansion. Its effect on the resulting LCs is to delay the onset of the flare rise and to smooth out any sharp features at the light-crossing timescale.

\section{Characteristics of simulated multiwavelength light curves}
\label{sec::lightcurves}

Three main scenarios have been studied: injection of particles following a power-law distribution (scenario 1),
Fermi-I acceleration (scenario 2), and Fermi-II acceleration (scenario 3). For the particle injection scenario, in addition to the basic scenario (1a) of injection, escape and radiative cooling, we consider also the impact of adiabatic expansion (scenario 1b). For the two acceleration scenarios, we consider first acceleration of an initially cold or mildly relativistic electron population (scenarios 2a and 3a), and then the reacceleration of relativistic electrons (scenarios 2b and 3b).

The physical parameters of the emission region remain unchanged, except in the case of adiabatic expansion, and are based on a flare study of the high-frequency peaked BL Lac object Mrk\,421 by \cite{Dmytriiev2021}. The same is true for the continuous particle injection spectrum responsible for the low-state emission in all scenarios. Mrk\,421 was chosen as a benchmark because it is an archetypal BL Lac object and is among the best-studied objects in this class, with some of the richest and most well-sampled multi-band data sets that can be tested against theoretical models. The parameter values adopted here are therefore seen as a representative baseline for high-frequency-peaked BL Lac objects.
Unlike in \cite{Dmytriiev2021}, where the EMBLEM code was applied in a model with several zones, evoking a combination of time-variable injection and stochastic reacceleration to reproduce the multi-band light curves of a specific flare from Mrk\,421, our aim here is not source-specific modeling but rather the extraction of generic flare signatures for different generic processes that should apply broadly to BL Lac objects with comparable parameter ranges.

For all simulations, the flaring phase lasts for a duration $t_{\rm dur} = 3$ observer days $\approx 8$ R/c. Outside of the flaring phase and during the entire simulation, when there is no particle acceleration, the escape timescale is fixed at $t_{\rm esc}=1$ R/c (cf. Table \ref{table_parameters}). The integral flux evolution has been studied in four different energy bands (cf. Fig.~\ref{appfig:sed_scenario1a}): VHE range ($2 \times 10^{11}$\hspace{0.1cm}-\hspace{0.1cm}$10^{14}$\,eV), HE range ($10^{8}$\hspace{0.1cm}-\hspace{0.1cm}$10^{11}$\,eV), X-rays ($2 \times 10^{2}$\hspace{0.1cm}-\hspace{0.1cm}$10^{4}$\,eV) and optical band ($1.5\hspace{0.1cm}-\hspace{0.1cm}3.5$\,eV).

\begin{figure}
    \centering
    \includegraphics[trim={0.5 0.5 0.5 1.5cm},clip,width=0.9\columnwidth]{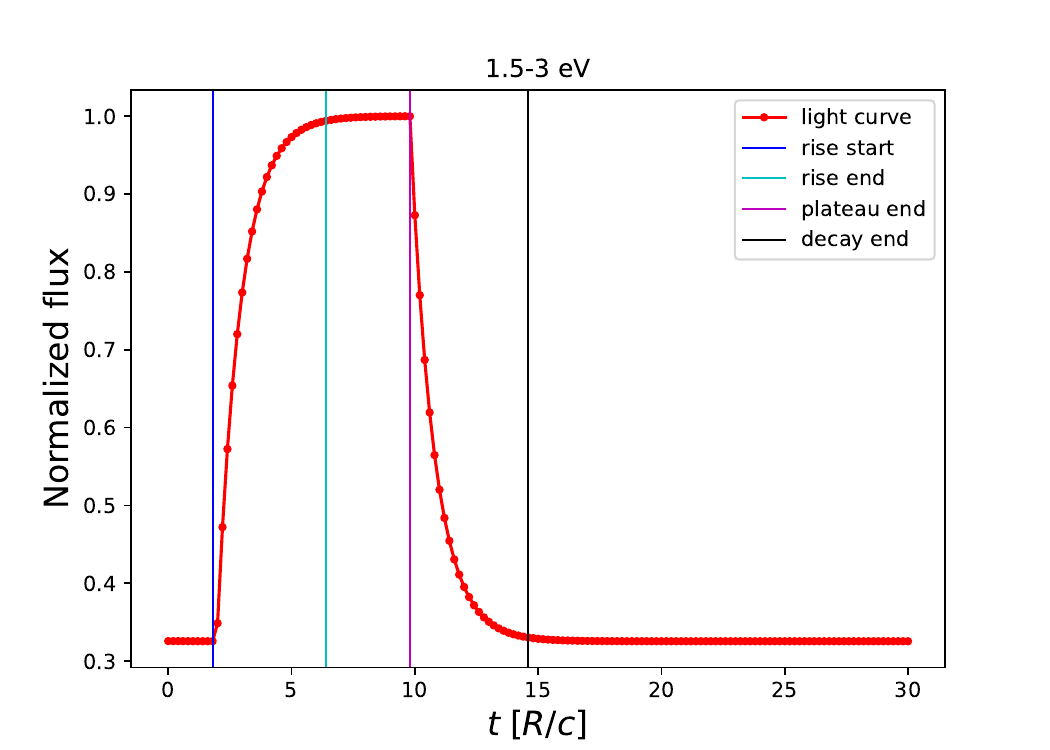}
    \caption{Example of LC obtained for the case of simple particle injection in the optical band, with the determined times of rise, plateau, and decay.}
    \label{LC_injection_times}
\end{figure}

LCs are simulated with a time step of 0.20 R/c. With $F(t_i)$ the normalized flux at time $t_i$,
the start of the flare $t_{\rm rise}^{\rm start}$ is determined such that $F(t_{\rm rise}^{\rm start}) - F(t_{\rm rise+1}^{\rm start}) > 0.001$. 
The end of rise $t_{\rm rise}^{\rm end}$ satisfies $F(t_{\rm rise+1}^{\rm end}) - F(t_{\rm rise}^{\rm end}) < 0.001$. 
In certain cases, a new steady-state appears during the flare, forming a plateau in the LC between $t_{\rm plateau}^{\rm start}=t_{\rm rise}^{\rm end}$ and $t_{\rm plateau}^{\rm end}$, the last time
step before the flare decay (cf. Figure~\ref{LC_injection_times}). The latter is defined by $t_{\rm decay}^{\rm start}$ and $t_{decay}^{\rm end}$ in a similar way to the rise. 
The durations of the rise ($\Delta t_{\rm rise}$), plateau ($\Delta t_{\rm plateau}$) and decay ($\Delta t_{\rm decay}$) are determined
from the start and end times. The asymmetry of the flare is determined as $k = (\Delta t_{\rm rise}-\Delta t_{\rm decay})/(\Delta t_{\rm rise}+\Delta t_{\rm decay})$ and the variability amplitude (VA) is given by the relative flux increase during the flaring event:
$A_{\rm var}=F_{\rm max}/F(t_{\rm rise}^{\rm start})$ with $F_{\rm max}=1$ the maximum normalized flux. 

\subsection{Scenario 1a: Injection}
\label{subsec::char1a}

The first scenario describes the occurrence of a flare caused by the injection of additional particles over a short duration. Physically, the injection of a power-law distribution could represent a case where a particle population is accelerated in a region with low magnetic field strength, which emits thus very little radiation, and is then injected into the emission region. A leading bow shock in front of a moving plasma blob, as suggested by \cite{Dmytriiev2021}, is a possible example. To explain the occurrence of flares, the injection rate needs to increase from the rate that characterizes the quiescent state, possibly due to inhomogeneities in the density of the plasma through which the shock is traveling.

To limit the number of free parameters to a minimum, a constant injection rate $\dot{N}_{\rm add}$ is assumed for a power-law particle distribution with the same spectral shape as the low-state injection, with $\gamma_{\rm add,min} =800$, $\gamma_{\rm add,pivot} =1.0 \times 10^5$, 
$\gamma_{\rm add,cut} =5.8 \times 10^5$ and index $\alpha_{\rm add} =-2.23$. For a systematic study, we vary the normalization $\dot{N}_{\rm add}=2.0 - 5.0\times 10^{-14}$ cm$^{-3}$s$^{-1}$ during a duration of $t_{\rm dur}$. Once the additional injection stops, the particle population will gradually reach its initial steady state through radiative cooling and particle escape with a constant escape time $t_{\rm esc}=1$\,R/c.

Even in this most basic scenario, the temporal evolution of the electron density cannot be described by a simple analytical expression due to the presence of radiative cooling, motivating our numerical approach. Initially, the electron distribution changes only in amplitude as a result of the additional injection. During the plateau and decay phase, the maximum electron energy
decreases due to cooling, until the initial steady state is reached. The electron distribution, for this and all subsequent scenarios, can be found on Zenodo\footnote{\url{https://doi.org/10.5281/zenodo.18430257}}.

The broad-band SED (Fig.~\ref{appfig:sed_scenario1a}) shows the expected increase in normalization during the flare. 
At the peak level, a small shift of both peaks to lower energies occurs because of cooling. This leads to a visible hysteresis, which is more pronounced for the synchrotron peak. The Compton dominance (CD), i.e.,\ the ratio between the SSC and synchrotron peak fluxes, remains inferior to 1, even for an increase in the SSC flux of an order of magnitude.

The LCs in the different energy bands and for different injection rates (cf. additional material on Zenodo) are very similar. The LC in Fig \ref{LC_injection_times} shows an example in the optical band.
The onset of the flare rise is the same in every band, as additional electrons are being injected at all energies, and the flux increase presents a concave shape. A plateau is reached when the processes of injection and loss balance out to reach a new steady state, given a sufficiently long flare duration. Radiative cooling leads to a migration of high-energy
electrons, responsible for the X-ray emission, to lower-energy electrons, responsible for the optical emission, such that the optical flux and HE flux
continue increasing slightly during the plateau phase, while the X-ray flux is
slightly decreasing. In the latter, this effect leads to a visible bump at the beginning of the plateau phase for sufficiently rapid injection. 

Apart from the slight change in shape during the plateau phase, the overall LC appears rather symmetric, but this depends on our assumptions of the additional injection rate and escape timescale.
For a sufficiently low rate of the additional injection, the asymmetry of the flare profile can be positive, i.e.,\ the rise time is longer than the decay time. LCs in the (V)HE bands present the strongest VA since the SSC bump rises more than linearly with $N_e(t)$ during the flaring phase.
The flux decrease after the flare is more rapid in the X-ray and VHE bands than in the optical and HE bands, because of the faster radiative cooling of the underlying electrons.

\subsection{Scenario 1b: Injection and adiabatic expansion}
\label{subsec::char1b}

This variant of the injection scenario includes the effect of adiabatic expansion of the blob for an assumed intrinsic opening angle $\alpha \approx 0.44\degree$, obtained by setting the viewing angle $\theta=1\degree$ and determining the bulk Lorentz factor of the blob for our given Doppler factor. The value of $\theta$ was chosen to be in agreement with the distribution of viewing angles estimated for the VLBI-detected blazars \citep[e.g.,][]{Hovatta2009, Savolainen2010, Liodakis2015}.

\begin{figure}
    \centering
    \vspace{-0.3cm}
    \includegraphics[trim={0.5 21 0.5 0.5cm},clip,width=0.95\columnwidth]{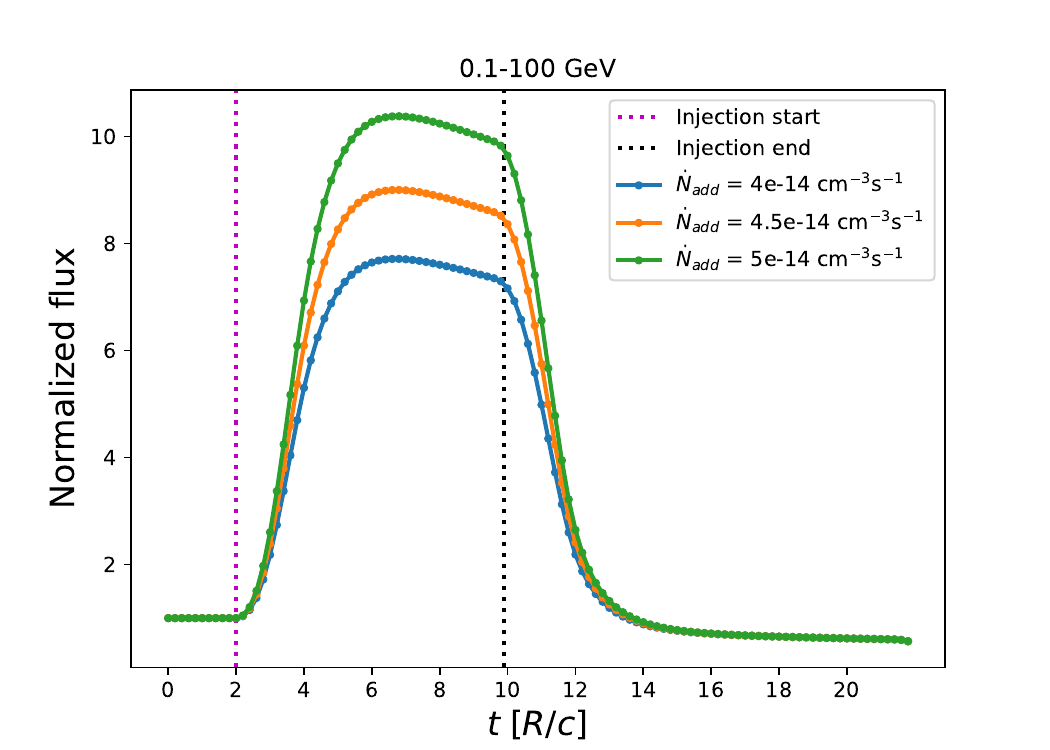}
    \includegraphics[trim={0.5 1.5 0.5 0.5cm},clip,width=0.95\columnwidth]{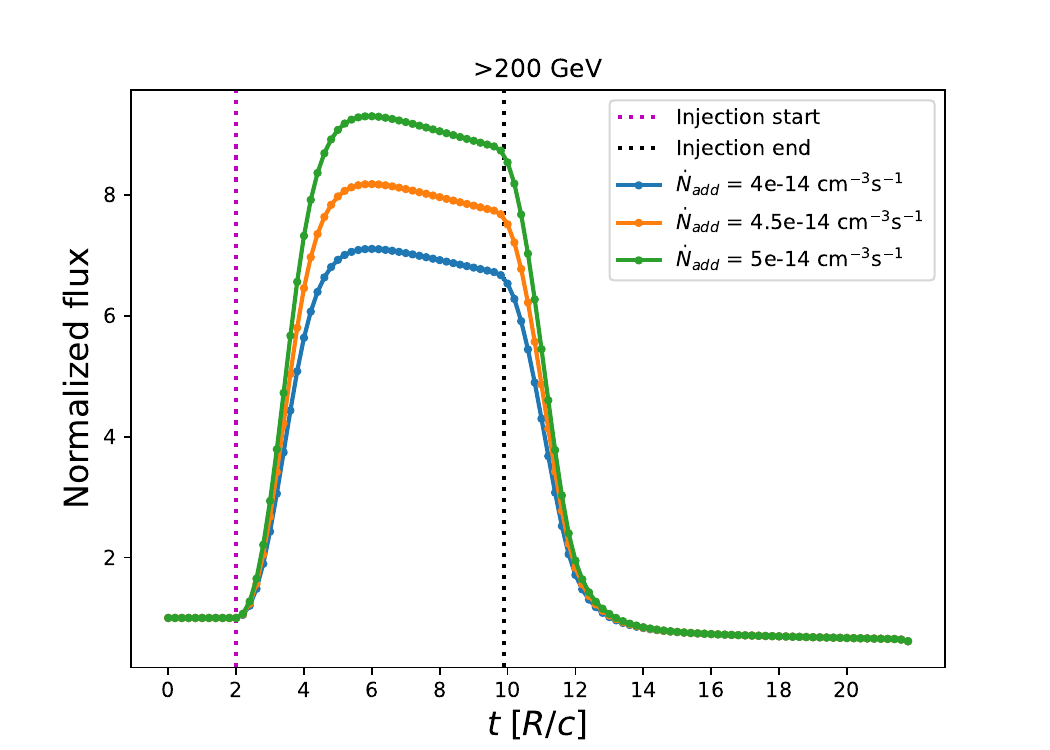}
    \caption{(V)HE LC of a flare resulting from additional particle injection with adiabatic expansion for different injection rates (scenario 1b).}
    \label{LCs_inject_adiab_compare}
\end{figure}

 As expected, the resulting SED is very similar to scenario 1a, with lower flux amplitudes due to the additional effects of particle cooling and the dilution of magnetic and particle density (cf. Fig.~\ref{appfig:sed_scenario1b}).
 The resulting (V)HE LCs are shown in Fig.~\ref{LCs_inject_adiab_compare} (time in units of the initial blob radius $R_0/c$) while the optical and X-ray LCs in Fig.~\ref{appfig:LCs_inject_adiab_compare}.
For LCs in all energy domains, including adiabatic expansion leads to decreasing steady-state emission during the low state and the plateau phase, inducing a negative asymmetry. 
It should be noted that the effect from adiabatic cooling is non-negligible for our standard parameters, even though the flare duration corresponds to only about three days in the observer frame.

\subsection{Scenario 2a: Fermi-I acceleration}
\label{subsec::char2a}

\begin{figure}[tb!]
    \centering
    \vspace{-0.3cm}
    \includegraphics[trim={0.5 21 0.5 0.5cm},clip,width=0.9\columnwidth]{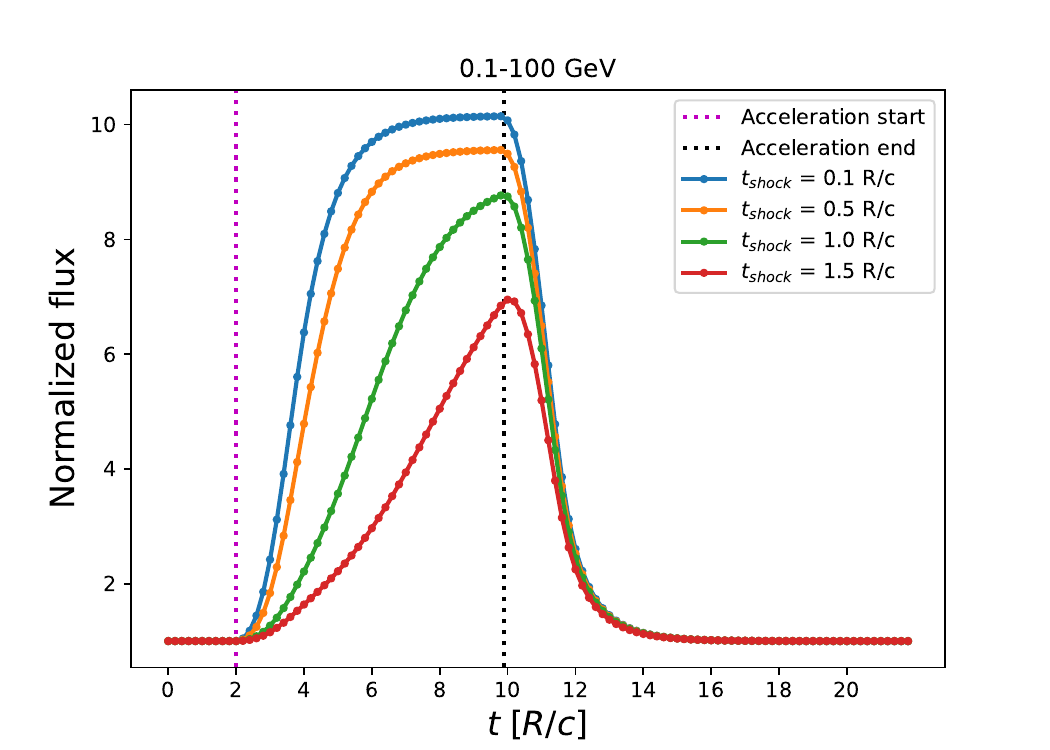}
    \includegraphics[trim={0.5 1.5 0.5 0.5cm},clip,width=0.9\columnwidth]{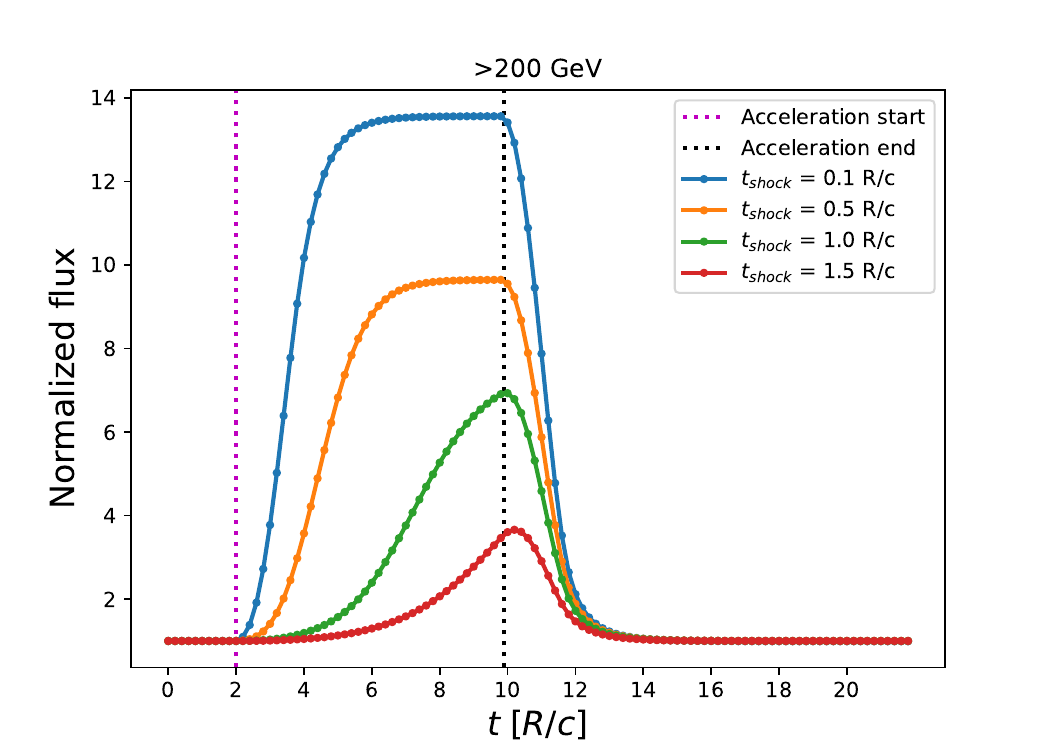}
    \caption{(V)HE LC comparison of flares resulting from DSA for different shock timescales (scenario 2a).}
    \label{LCs_FI_acc_compare}
\end{figure}

In the first acceleration scenario, the flaring is caused by DSA during $t_{\rm dur}$ with a shock timescale $t_{\rm shock}$ that
determines the rate of increase of $\gamma_{\rm max}$. This scenario represents the effects of a shock accelerating and injecting particles into the emission
region, due to the encounter of the region with a stationary or moving shock. The values of $\gamma_{\rm min}$ and the slope of the power-law distribution are the same as for the injection during the low state, but the cutoff Lorentz factor evolves during the flaring phase. Apart from $t_{\rm shock}$, the injection rate of the shock-accelerated particles $\dot{N}_{add}$ is the only other free parameter.
No expansion of the blob is shown here or in the following scenarios, so that the source parameters remain constant with time. The effect of adiabatic expansion on the LCs are similar to the case of simple particle injection. The escape timescale is fixed at $t_{\rm esc}= 1$ R/c. 

We present scenarios for flares resulting from DSA for a fixed $\dot{N}_{add}=4 \times 10^{-14}$\,cm$^{-3}$ and a range of shock timescales, from $t_{shock}= 0.1$ R/c corresponding to $\gamma_{max} \sim 5.2 \times 10^6$, to $t_{shock}= 1.5$ R/c corresponding to $\gamma_{max} \sim 3.5 \times 10^5$.
All other parameters are identical to those of Section \ref{subsec::char1a} (cf. Table~\ref{table_parameters}). 

During the flaring phase, the electron density of the additional injection increases with time for increasingly high energies (cf. additional material on Zenodo).
The SED translates this behavior with a flux increase and a shift of the peaks first to energies lower than the low-state peaks, and then to higher energies toward the end of the flaring phase (cf.~Fig~\ref{appfig:sed_scenario2a}). Once acceleration ends, the same reversed behavior is observed, although the shifts are less pronounced. The hysteresis for both peaks is more pronounced than in the injection scenario.

In the limit of very efficient acceleration, by design the simple injection scenario 1a is reproduced, as seen for the blue curve in Fig.~\ref{LCs_FI_acc_compare} and Fig.~\ref{appfig:LCs_FI_acc_compare}. 
For intermediate acceleration timescales (e.g.,\ $t_{\rm shock}=0.5$\, R/c, orange curve), the LC reaches a steady state at different times depending on the electron energy and thus the energy band. 
In this case, since the maximum Lorentz factor $\gamma_{\rm add,cut}$ increases with time during acceleration, the LC peak in the X-rays occurs later than in the optical band for $t_{\rm shock}\gtrsim 0.5$ R/c. The onset of the flux increase is visibly delayed between these bands for sufficiently long acceleration timescales. This behavior is reflected between the HE $\gamma$-ray and VHE $\gamma$-ray domains. 

In the X-ray band, before the onset of the flare rise, the flux amplitude is seen to slightly decrease, because of enhanced IC cooling on the increased synchrotron peak flux. For long shock timescales $t_{\rm shock} \gtrsim 0.8$ R/c (cf.\ green and red LCs in Figure~\ref{LCs_FI_acc_compare}), the flare does not reach the  plateau phase by the end of acceleration. The flux increases for a longer time than in the case of short shock timescales, since an additional population of electrons is continuously added at higher energies and cools down to lower energies, as $\gamma_{max}$ increases.

For all bands, the VA is roughly linear to the inverse shock timescale.
For moderate acceleration timescales $t_{shock} \lesssim 0.5 R/c$, the LCs are marked by a VA that is smaller in the X-rays and the VHE band than in the optical and the HE band. 
For all LCs, the rise shape is first convex and then concave if a steady state is approached. When $t_{\rm shock}$ increases, the concave part of the rise disappears. At $t_{\rm shock}\sim 0.1$ R/c, the asymmetry for all bands remains in the range $-0.04 \leq k \leq 0.06$. It increases as the shock timescale increases, because the rise time lasts longer and becomes larger than the decay time. For the maximum timescale considered, $t_{\rm shock}=1.5$ R/c, the asymmetry is between $0.19 \leq k \leq 0.33$.

\subsection{Scenario 2b: Fermi-I reacceleration}
\label{subsec::char2b}

\begin{figure}[tb!]
    \centering
    \includegraphics[trim={0.5 21 0.5 0.5cm},clip,width=0.9\columnwidth]{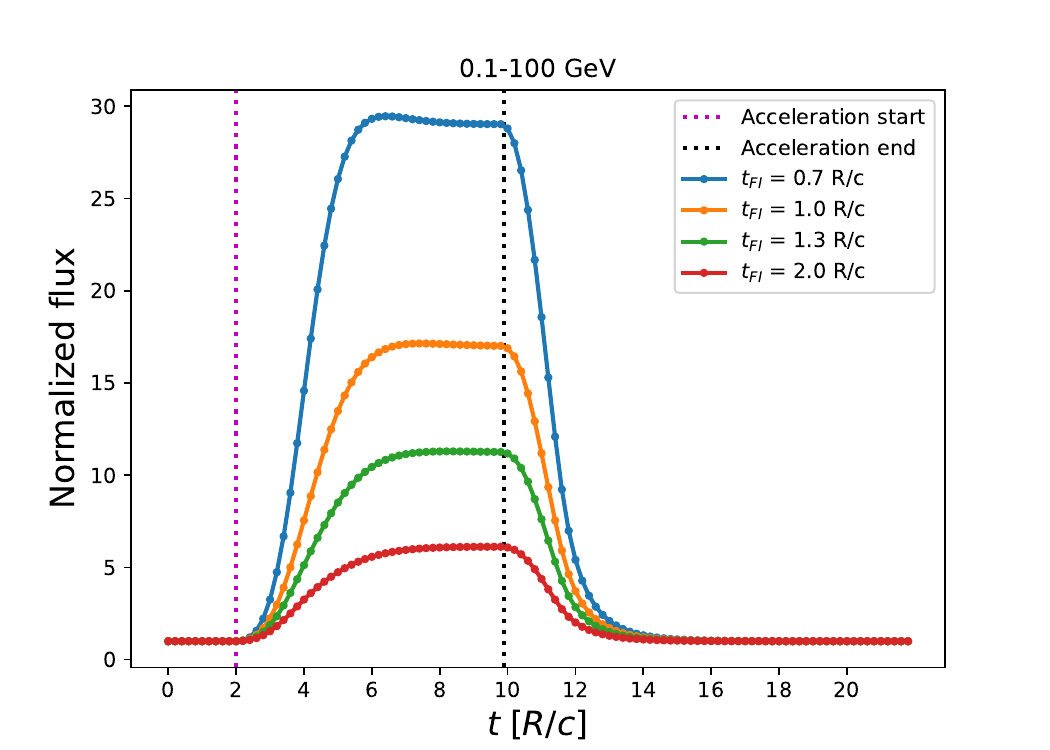}
    \includegraphics[trim={0.5 1.5 0.5 0.5cm},clip,width=0.9\columnwidth]{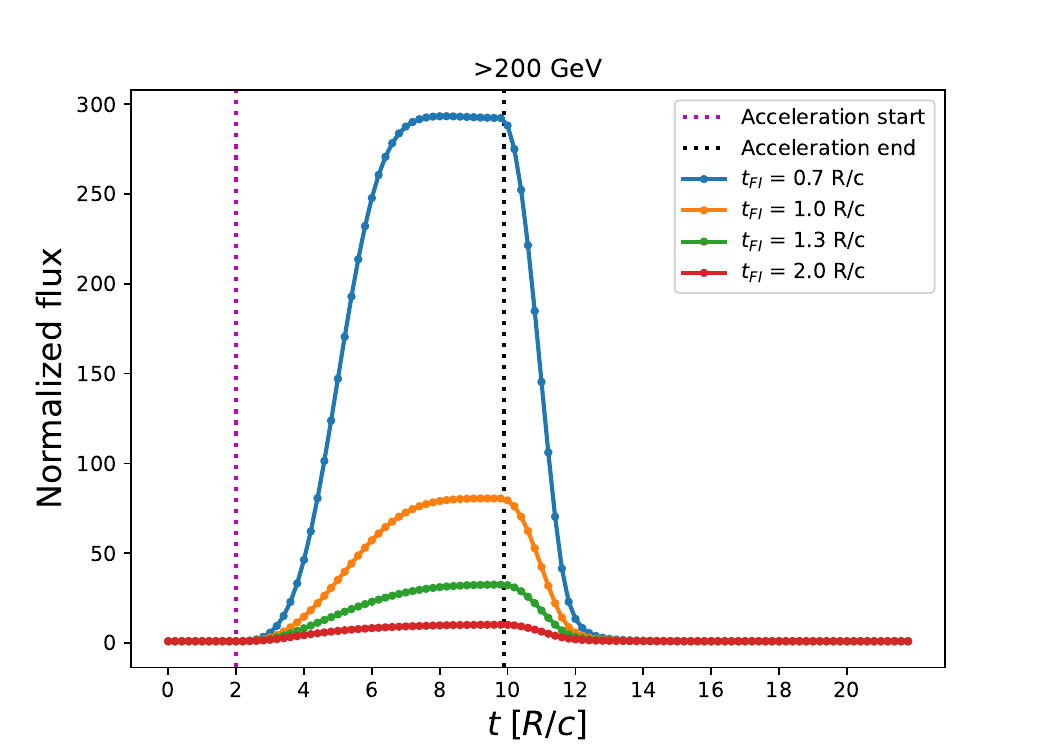}
    \caption{(V)HE LC comparison of flares resulting from Fermi I reacceleration for different timescales $t_{F_I}$ (scenario 2b).}
    \label{LCs_FI_reacc_compare}
\end{figure}

In addition to the above scenario, where flaring arises from a new population of electrons accelerated on a relativistic shock, we also consider the case where the relativistic electrons responsible for the quiescent-state emission are themselves reaccelerated on a shock inside the emission region. Depending on the particle density and energy distribution in the emission region, either scenario~2a or scenario~2b could be more appropriate to describe the actual
effect of insitu shock acceleration. Reacceleration scenarios are of particular interest to explain the hardest observed spectral distribution and have been applied for example to explain the steady emission of extreme blazars \citep[e.g.,][]{Zech2021,Tavecchio2022}.
Acceleration timescales are chosen between 0.7 and 2.0 R/c to cover a large range of VAs. The escape timescale is again set to $t_{\rm esc}= 1$ R/c.

In the two reacceleration scenarios we are treating (scenarios 2b and 3b), the particle injection does not differ from the one during the quiescent state, but particles gain additional energy, resulting in harder spectra.
This leads to a shift to higher energies of the SED peaks and to an increase of their amplitudes. The strong IC cooling of electrons in the flare decay phase causes a strong hysteresis in the synchrotron peak (cf. Fig.~\ref{appfig:sed_scenario2b}). For all acceleration timescales, the CD remains inferior or approximately equal to 1.

The optical and X-ray LCs (Figure~\ref{appfig:LCs_FI_reacc_compare}) show a marked difference in the shape of the rising flux, which is reflected in the HE and VHE $\gamma$-ray bands (Figure~\ref{LCs_FI_reacc_compare}). In the optical and HE bands, a high steady state is rapidly reached, reflecting an equilibrium between energy gain, cooling and escape. The continuous acceleration leads to a longer increase in the population of those electrons responsible for the X-ray emission, with the X-ray flux reaching its peak only when reacceleration stops. In this band, a small plateau is only visible for the shortest acceleration timescale considered here. The VHE LC shows a similar behavior.
In the optical band, for sufficiently short acceleration timescales, a bump occurs in the flux at the beginning of the plateau, similar to the signature in the X-ray band in the injection scenarios 1a and 1b.

The longer rise time in the X-ray and VHE bands leads to an asymmetric shape of the LCs, while they are nearly symmetric in the optical and HE band.
The VA is always largest in the (V)HE LCs, translating the fast rise of the SSC bump during the flaring phase. In particular, the VHE band has the strongest flux variation because the SSC peak is shifted to higher energies during the flaring and is located closer to this energy domain. The increase in emission visible in the SED is very concentrated around the synchrotron and SSC peaks, leading to a larger VA in the X-rays and VHE $\gamma$-rays compared to the optical and HE $\gamma$-ray bands, respectively.

\subsection{Scenario 3a: Fermi-II acceleration}
\label{subsec::char3a}

Alternatively, flaring may arise from stochastic acceleration of particles, with a time-dependent escape timescale $t_{\rm esc}^{\rm (turb)}$. Stochastic acceleration might be due to the emission region traversing a zone of turbulence inside the jet. 

We consider again only the hard-sphere turbulence scenario
and, given the degeneracy between $\lambda_{max}$ and $(B/\delta B)^2$ (cf. Equ.~\ref{equ:t_acc_fermi2}) in the absence of adiabatic expansion, only the perturbation level of the magnetic field was varied, leading to a range of timescales with minimum and maximum values of $t_{\rm F_{II}}^{\rm min} \sim 1.9$ R/c and $t_{\rm F_{II}}^{\rm max} \sim 4.1$ R/c. 
The particle population before acceleration is assumed to consist of cold or mildly relativistic electrons, which we represent for simplicity with a narrow logparabolic distribution: 
$\dot{Q}_{inj} = 10^{-8} (\gamma/10)^{-6.0 -4.0 \log(\gamma/10)}$ cm$^{-3}$s$^{-1}$.
Any emission from the initial particle distribution is negligible. We assume this population to occupy the same zone as the relativistic electrons responsible for the low-state emission, so the energy density of the latter is taken into account when determining the Alfvén speed\footnote{For simplicity, the relativistic formulation is applied to all particles.} that influences the acceleration and escape timescales.  Reacceleration of the relativistic electrons is assumed to be negligible and the particle density of the cold electrons is chosen such as to be an order of magnitude higher than the relativistic ones.

As stochastic acceleration sets in, the initial logparabolic particle distribution sees its peak shift to higher energies and develops a hard power-law tail with increasing cutoff energy. 
As acceleration stops, radiative cooling shifts the peak back to lower energies and the high-energy electrons escape from the source. The SED (Fig.~\ref{appfig:sed_scenario3a}) is the sum of the low-state and flaring emission. It is marked by a synchrotron peak energy that increases during the acceleration phase, but does not reach the value of the low-state peak in the given example. The effect of hysteresis is small in this scenario since the peak energies do not change significantly.

\begin{figure}[tb!]
    \centering
    \vspace{-0.02cm}
    \includegraphics[trim={0.5 21 0.5 0.5cm},clip,width=0.9\columnwidth]{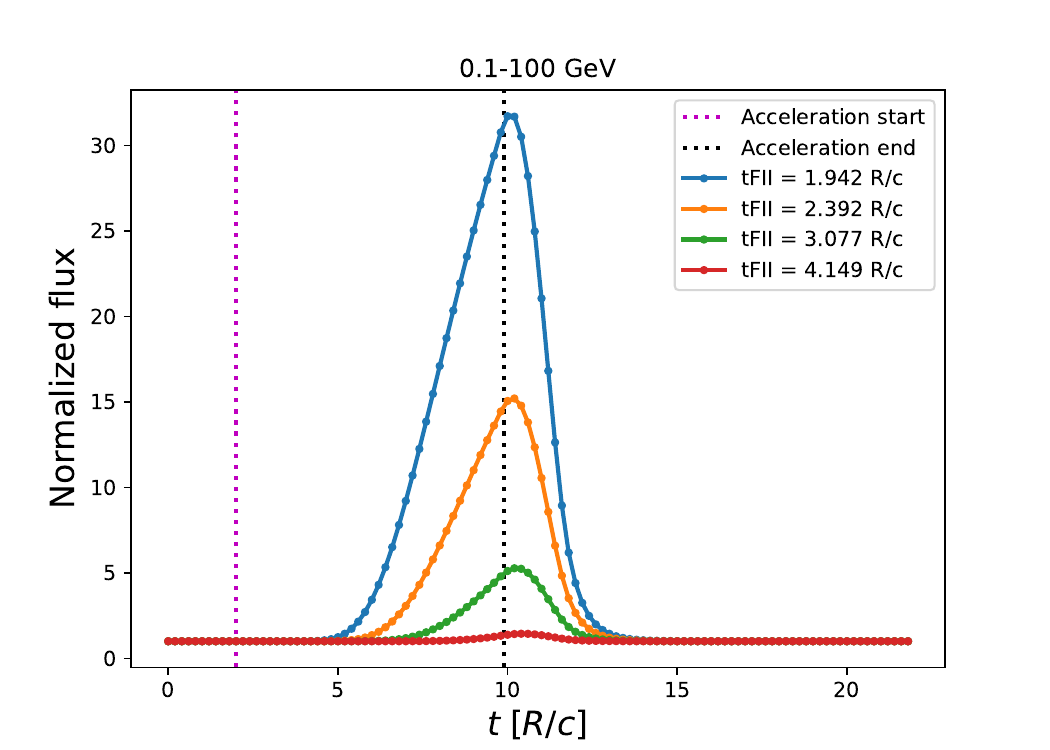}
    \includegraphics[trim={0.5 1.5 0.5 0.5cm},clip,width=0.9\columnwidth]{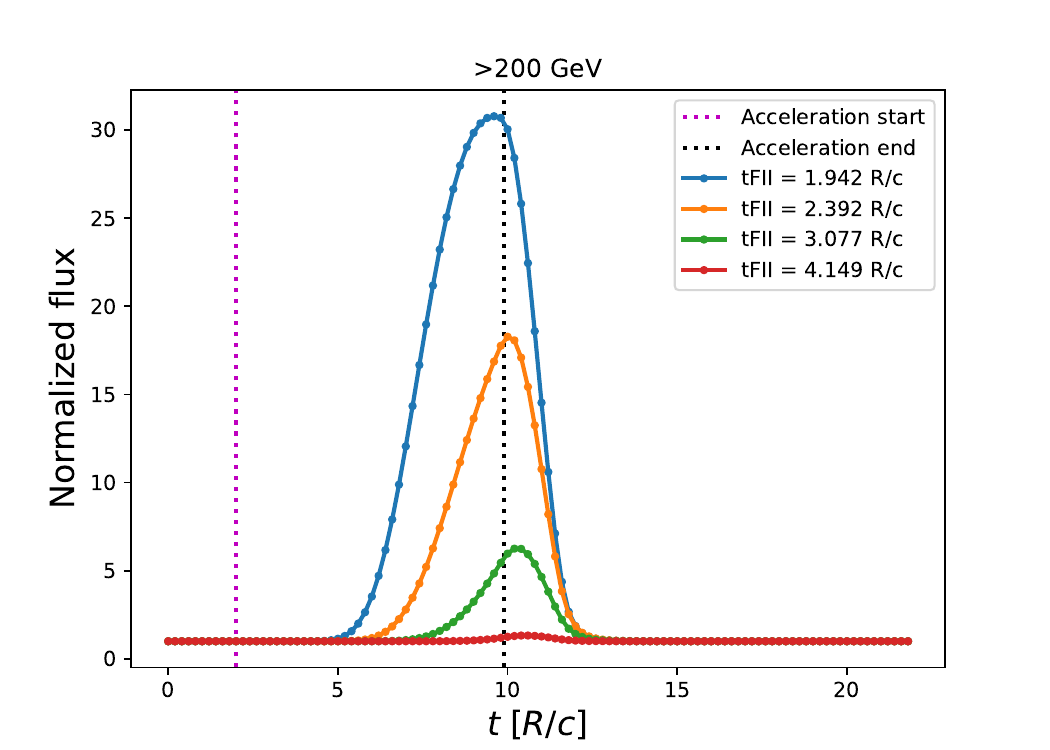}
    \caption{(V)HE LC comparison of flares resulting from hard-sphere Fermi II acceleration for different acceleration timescales $t_{F_{II}}$ (scenario 3a).}
    \label{LCs_FII_acc_compare}
\end{figure}

The LCs are shown in Fig.~\ref{LCs_FII_acc_compare} and Fig.~\ref{appfig:LCs_FII_acc_compare} for different acceleration timescales. The legend indicates the initial acceleration timescale for each simulation, keeping in mind that this timescale evolves during the flaring phase. 
As in scenario 2a, the onset of the flux rise depends on the acceleration timescale, but it does not change markedly between energy bands.
In contrast to scenarios 1 and 2, the LCs do not reach a steady state during flaring for the wide range of VA explored here, due to the continuous systematic energy increase and broadening of the particle distribution during the acceleration phase. The rising part of the LC is always convex. The peak amplitude is generally reached after the acceleration has already stopped, as a consequence of the light-crossing effect, which is particularly visible for this configuration due to the sharp peaks in the LCs. An exception to this is the LC for the
shortest acceleration timescale in the X-ray band, where IC cooling stops the increase of the flux before the end of acceleration is reached.

While the escape timescale during the flaring activity depends on the acceleration timescale, once the flare stops particle escape is again described with a timescale of 1\,R/c. Thus, the falling slope of the LC depends only on the usual escape and cooling losses.
The overall flare shapes are close to symmetric in all bands.

\subsection{Scenario 3b: Fermi-II reacceleration}
\label{subsec::char3b}

As for the Fermi~I case, we again consider a second Fermi-II scenario assuming reacceleration of the power-law distribution of the relativistic particles responsible for the quiescent-state emission, without additional injection. 
The maximum wavelength and turbulence level are varied to span an order of magnitude in the acceleration timescale, with minimum and maximum timescales $t_{\rm F_{II}}^{\rm min} \sim 1.4$ R/c and $t_{\rm F_{II}}^{\rm max} \sim 14.9$ R/c.

 Initially, the particle distribution hardens, leading to an increase of the synchrotron flux and to a rapid increase of the SSC peak flux and CD above 1 (cf. Fig.~\ref{appfig:sed_scenario3b}). The most energetic electrons then rapidly lose their energy to IC cooling, and the synchrotron peak shifts to energies well below the steady-state peak energy, while the SSC peak flux decreases. The positions of both peaks describe a clear hysteresis in the space of peak flux versus peak frequency.

\begin{figure}[tb!]
    \centering
    \includegraphics[trim={0.5 21 0.5 0.5cm},clip,width=0.9\columnwidth]{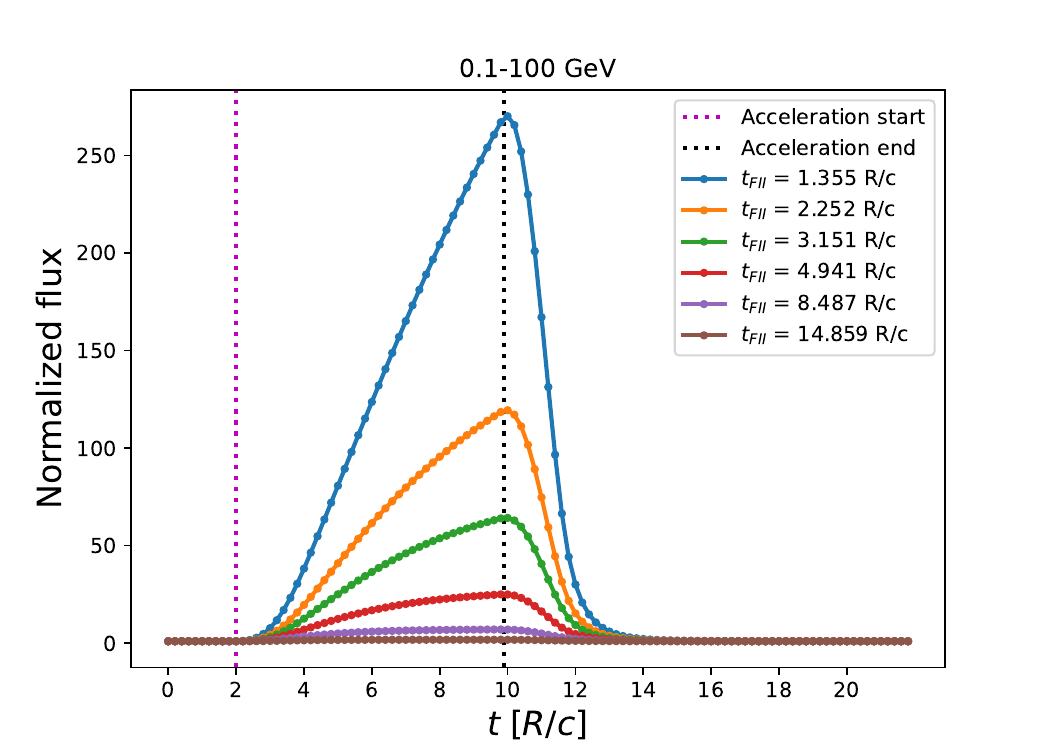}
    \includegraphics[trim={0.5 1.5 0.5 0.5cm},clip,width=0.9\columnwidth]{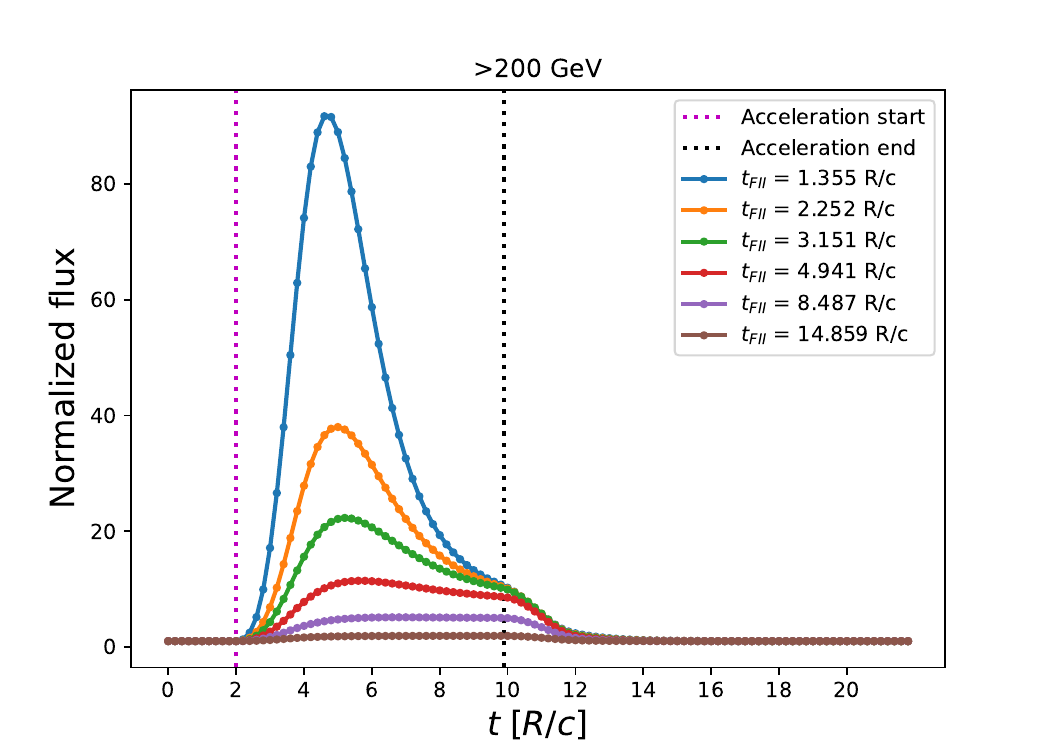}
    \caption{(V)HE LC comparison of flares resulting from hard-sphere Fermi II reacceleration for different timescales $t_{F_{II}}$ (scenario 3b).}
    \label{LCs_FII_reacc_compare}
\end{figure}

The LCs of these simulations are plotted in Fig.~\ref{LCs_FII_reacc_compare} and Fig.~\ref{appfig:LCs_FII_reacc_compare}, with the legend indicating the initial acceleration timescale for each simulation.  
The VA of the flares steadily increases for decreasing acceleration timescales. In the X-ray and VHE $\gamma$-ray bands, the flare peak shifts to an earlier time. The more efficient acceleration implies that the flux quickly increases in these bands, but this increase is quenched by IC cooling as the flaring phase is ongoing. Such flares present a CD well above 1, so that they correspond to what we define as the "high CD regime".

As in scenario 3a, the LCs do not reach a plateau phase, except for the most inefficient cases with low VA. In the high CD regime, the rising part of the flare is markedly different between the long concave rise in the optical and the very rapid increase in the X-ray band, both reflected in the HE and VHE bands. The time delay between the peaks in the two sets of bands is a very visible marker of this scenario. 

The flux decrease begins already during the flaring phase in the X-ray and VHE bands and is significantly slower than in the optical and HE bands. As a result, the asymmetry of the flares in the optical and HE bands is always positive while it is negative in the X-ray and VHE ones.

\section{Discussion}
\label{sec::discussion}

\subsection{Comparison of observable features}
\begin{figure*} 
    \centering
    \includegraphics[trim={0.5 21 0.5 0.5cm},clip,width=0.9\columnwidth]{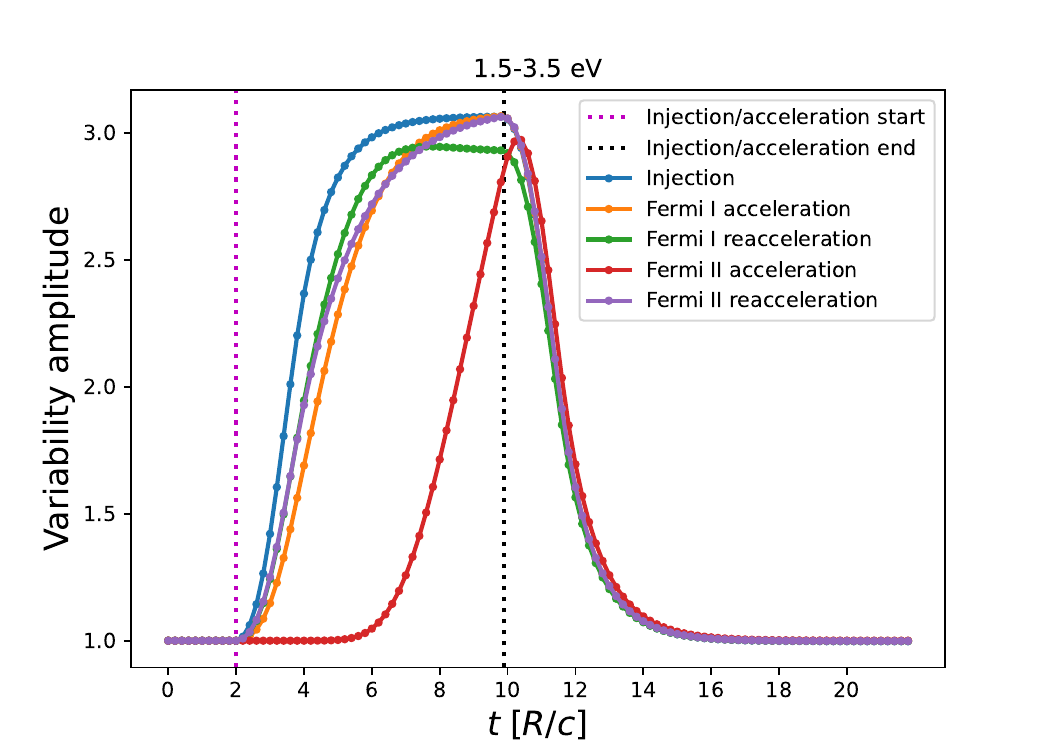}
    \includegraphics[trim={0.5 21 0.5 0.5cm},clip,width=0.9\columnwidth]{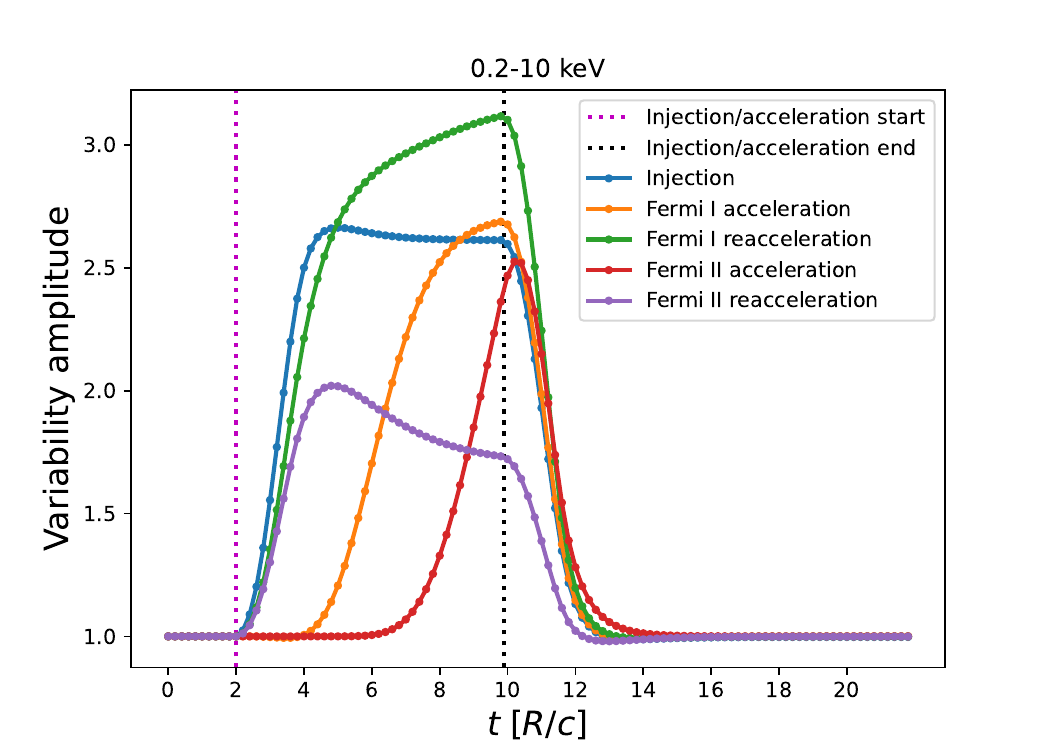}
    \includegraphics[trim={0.5 1.5 0.5 0.5cm},clip,width=0.9\columnwidth]{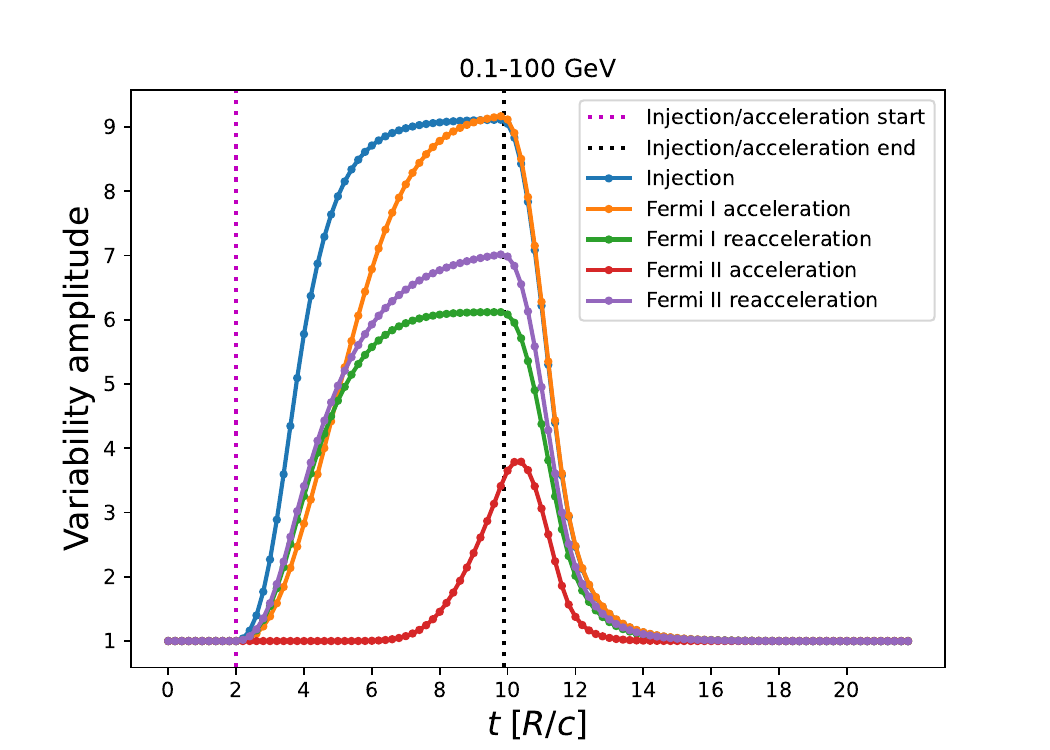}
    \includegraphics[trim={0.5 1.5 0.5 0.5cm},clip,width=0.9\columnwidth]{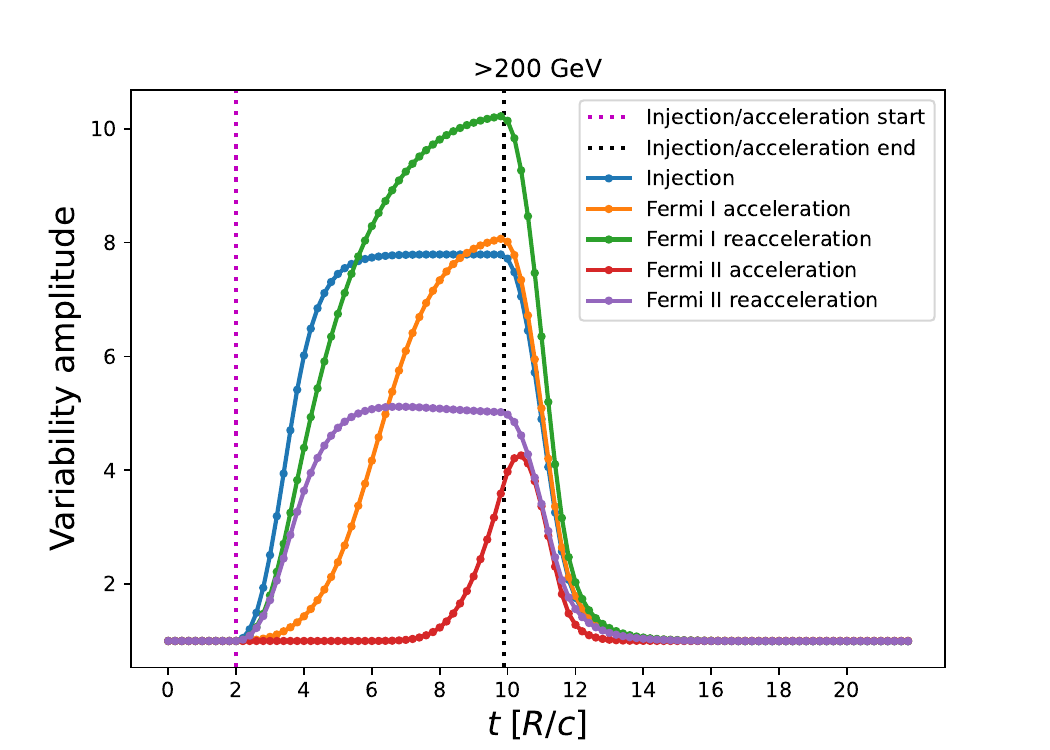}
    \caption{LC comparison of the flaring scenarios corresponding to particle injection (1a), Fermi-I acceleration (2a), Fermi-I reacceleration (2b), Fermi-II acceleration (3a), and Fermi-II reacceleration (3b).}
    \label{LCs_scenarios_compare}
\end{figure*}

The comparison of LCs between different scenarios shows a few characteristic features that may be exploited observationally. For a direct comparison, the LCs of the acceleration and injection scenarios without adiabatic expansion are shown in Fig.~\ref{LCs_scenarios_compare}, using parameters for which the VA $\sim 3$ in the optical band. 

For the same VA in the optical band, the different scenarios result in a broad range of VAs in the other bands. While similar levels are seen for the closely related injection and Fermi-I scenarios, the Fermi-II VA is markedly lower in the HE and VHE bands (SSC component). 
The reacceleration scenarios show similar behavior in the optical and HE bands, associated with electrons of lower energy, but a large difference in the X-ray and VHE bands, which are dominated by the most energetic electrons. 

The onset of the flare rise occurs approximately at the same time in all bands for the injection and for the reacceleration scenarios, as can also be seen from Fig.~\ref{LCs_inject_adiab_compare} to~\ref{LCs_FII_reacc_compare} (and Fig.~\ref{appfig:LCs_inject_adiab_compare} to~\ref{appfig:LCs_FII_reacc_compare}). In the Fermi-I initial acceleration scenario 2a, however, the flare rise begins at significantly different times across bands, with the higher energy bands showing a delayed response. A small delay is also visible for the Fermi-II acceleration scenario 3a.

No delays between the LC peaks in different bands are seen for low VAs in the injection and Fermi-I scenarios, but efficient acceleration or rapid injection, leading to high VAs, induces a small peak in the X-ray band before entering the high-state plateau. In scenario 3b, and for the highest VAs in scenario 3a, the peak in the X-ray band is reached significantly earlier than in the optical band. For the latter, this delay is not reflected in the VHE band. A particularity of the Fermi-II reacceleration scenario 3b is the notable difference in LC shapes between the optical and X-ray bands, as well as between the GeV and TeV bands.

A plateau in the flare LC can be reached in the injection scenarios for a large range of VA values, if the flare duration is sufficiently long. As expected, scenario 1b exhibits a systematic decrease of the flux during the plateau phase in all bands. In scenario 2a, the plateau phase is only reached for very efficient acceleration, i.e.,\ very high VAs, while in scenario 2b, it can be reached for a large range of VAs. Scenario 3a does not exhibit a plateau phase, while in scenario 3b a plateau can be reached for low VAs, i.e.,\ inefficient acceleration.
The flare rise time shows little variation for different VAs in the injection scenarios, whereas in the Fermi-I and Fermi-II (re)acceleration scenarios, it is shorter for higher VAs, as indicated by the differing slopes of the flare rise.

Judging from the SEDs (Fig.~\ref{appfig:sed_scenario1a} to~\ref{appfig:sed_scenario3b}), the injection scenarios present the least hysteresis, since we do not have a significant shift in the peak positions.
Hysteresis is seen for both peaks in the Fermi-I and Fermi-II (re)acceleration cases, where the peaks shift to higher energies during the flare rise and to lower energies during the cooling. The hysteresis is most pronounced in the Fermi II reacceleration scenario, especially for the synchrotron bump in the case of short reacceleration timescales.

Based on the signatures discussed above, it is clear that well sampled LCs from at least two energy bands are necessary to disentangle the scenarios, even within a single-zone SSC model framework. In particular, comparing LCs that sample both the synchrotron and SSC bumps is essential. Access to multiple well-sampled energy bands allows us to capture the behavior of a wider portion of the electron population and relate its evolution to a flaring scenario. Comparing the onset of flare rise and peak times across different energy bands appears to be the most promising way to assess the underlying injection or acceleration mechanism. 

\subsection{Exploration of systematic effects}
As a variant of scenario 2a, we have investigated the effects of energy-dependent Fermi I-acceleration with one example. Following the approach described in the Appendix A of \citet[][]{Kirk1998}, we assume the acceleration and escape timescales to scale linearly with energy. In this scenario, the electron distribution and SED change significantly depending on the initial timescales $t_{shock,0}$ and $t_{esc,0}$. For values in the range $10^{-6}-10^{-4}$ R/c, allowing to obtain an electron distribution peaking between Lorentz factors of $10^4-10^5$ during the flaring phase, the SED presents a marked hysteresis in the synchrotron peak and a weaker hysteresis for the SSC one. In the X-ray and VHE bands, the LCs reach a plateau during the acceleration phase, which leads to smaller asymmetries than in the optical and HE bands. The difference in rise times between bands is not evident as it is in the energy-independent scenario but the faster decay in the X-ray (VHE) band compared to the optical (HE) one remains visible. A full description of energy-dependent acceleration scenarios is left for future multi-zone modeling.

We have also explored the case of Fermi-II reacceleration assuming a Kolmogorov turbulence spectrum with $q=5/3$ and a Kraichnan turbulence spectrum with $q=3/2$ \citep[e.g.,][]{Asano2014}. In these cases, the escape and acceleration timescales become energy-dependent (cf. Equ.~\ref{equ:tesc_fermi2} and~\ref{equ:t_acc_fermi2}).   
For physically relevant intervals of the maximum wavelength of the turbulent spectrum $0.1\leq \lambda_{max}\leq 1$ and of the turbulence level $0.01\leq \mathcal{L}_{turb} \leq 1$, the Kolmogorov and Kraichnan acceleration timescales remain below 1\,R/c. The resulting LCs therefore have the same shapes as the high CD hard-sphere flares of scenario 3b, in all energy bands. In general, one finds very
similar LC shapes between the three turbulence regimes, but for different values of $\lambda_{max}$ and $\mathcal{L}_{turb}$. 

In the Thomson regime, the strong Compton dominance we find in the reacceleration scenario 3b would lead to a nonlinear cooling process, quenching the energetic electron population in a "Compton catastrophe" \citep[e.g.,][]{Petropoulou2015}. However, given our chosen set of parameters corresponding to a high-frequency peaked BL Lac object, this effect is countered by the
rapid decrease of the cross-section in the Klein-Nishina regime. \citet{Petropoulou2015} provide a simple expression for the highest order of up-scattered photons in the Thomson regime, in the case of a mono-energetic electron distribution at Lorentz factor $\gamma_0$:
\begin{equation}
N_T=\left[ \frac{log(B_{cr}/B\gamma_0)}{log(4\gamma_0^2/3)} \right],
\end{equation}
with $B_{cr}= m_e^2 c^3 / (e \hbar) \approx 4.4 \times 10^{13}$\,G. 
For our chosen parameters, setting $\gamma_0$ to our minimum Lorentz factor, we find $N_T = 2$, indicating that secondary SSC might still need to be considered in our "high CD regime." This effect has thus been verified
for one of the most efficient reacceleration scenarios, and we find that the
contribution of secondary SSC is several orders of magnitude below the primary SSC emission.

While a specific set of source parameters was used for the models presented here, the evolution of the electron distribution and the SED show a qualitatively similar behavior if the parameter set is changed.
The shapes of the extracted LCs in fixed energy bands naturally depend on the relative position of those bands with respect to the SED peak positions.
We have verified the effect of reducing the magnetic field to $B=0.01$\,G  and the source extension to $R=1 \times 10^{16}$\,cm (cf. additional material on Zenodo).
In the case of the smaller blob size, the duration of the flare had to be reduced to one day in the observer frame, to keep the same observed flare duration. 

The relative VAs, flare onsets, flare shapes etc.\ remain consistent in both cases for all scenarios, except the Fermi-II ones, for changes of $R$ and $B$ within an order of magnitude. In the case of Fermi II acceleration, the temporal properties of the flares remain the same, but the relative VA changes between energy bands. For a smaller magnetic field, the VA becomes stronger in the optical compared to the X-ray and in the HE compared to the VHE, while for a smaller blob radius the opposite effect is observed. In the case of Fermi-II reacceleration, the differences in relative VA and shape arise from the significant changes in the LCs with the acceleration timescale, as discussed in Section~\ref{subsec::char3b}.

\section{Comparison to a 2013 flare of Mrk\,421}
\label{sec::application}

\begin{figure}[htb!]
    \centering
    \includegraphics[trim={0.5 14 0.5 0.5cm},clip,width=0.9\columnwidth]{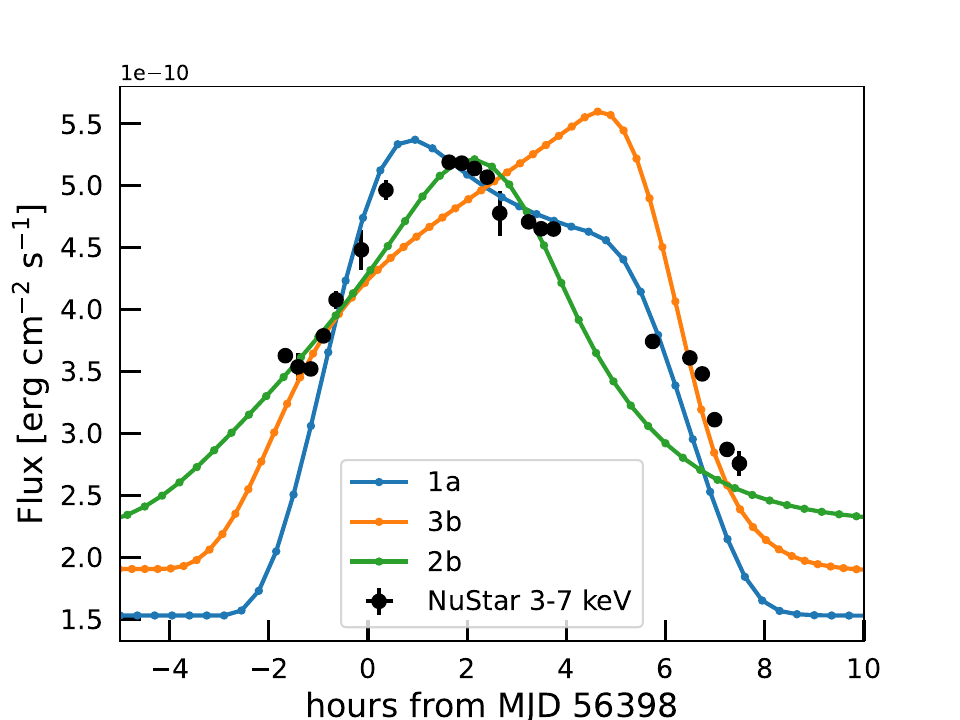}
    \includegraphics[trim={0.5 0.5 0.5 1cm},clip,width=0.9\columnwidth]{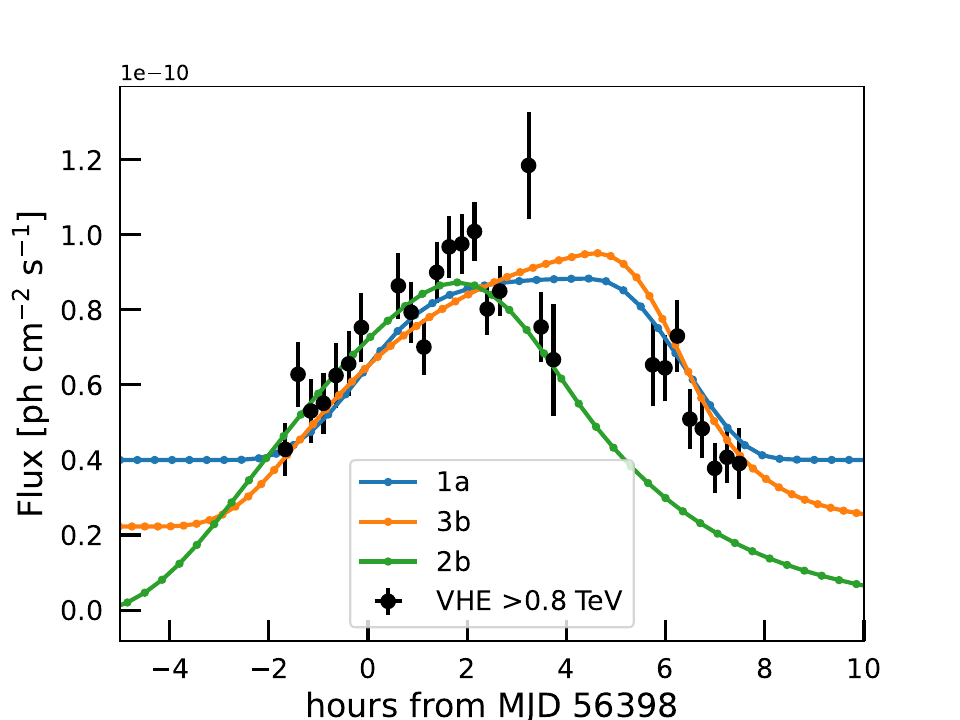}
    \caption{Comparison of selected flare scenarios to a set of X-ray and VHE flare LCs from the blazar Mrk\,421. Data points are taken from \citet{Acciari2020} }
    \label{LCs_application_Mrk421}
\end{figure}

To illustrate the interest, as well as the difficulties, of applying the different scenarios to observational flare data sets, we compare an injection, Fermi-I reacceleration and Fermi-II reacceleration scenario to a data set taken in the X-ray and VHE range during a rapid flare of the blazar Mrk\,421 on April 16th, 2013 \citep{Acciari2020} in Fig.~\ref{LCs_application_Mrk421}. These three scenarios were chosen since they are best suited to reproduce flares which are most dominant at the highest energies. This first application does not represent a detailed modeling, which would require a reproduction of the full time-resolved SED. As was already seen for the 2010 flare \citep{Dmytriiev2021}, a single emission region cannot account for the full multiwavelength emission of the source during the flare. A compact emission region is evoked here to represent the well sampled variable high-energy emission, while the continuous broad-band emission at lower energies is ascribed to a different unspecified zone. Since the baseline of the high-energy flux before and after the flaring event is not known, we introduce two free parameters that correspond to the normalization of the baseline flux in the X-ray and VHE band.

For all three scenarios, the parameters of the compact emission region were adjusted to approximate the flux variation in the X-ray and VHE band. Our choice for this study to not introduce any arbitrary time-dependent variations of the additional injection level or spectral shape during flaring puts strong constraints on the attainable LC shapes. The need to reproduce the rise and fall times constrains the timescale associated with one light-crossing time in the observer frame and thus the ratio of ${R_b}/{\delta_b}$. The fall time also constrains the escape timescale, which was set to values of $5-8$ R/c here. 

Compared to the input parameters used for the systematic studies, the emission region for this application requires a smaller radius and higher magnetic field, while its initial electron spectrum is shifted to higher $\gamma_{min}$ and $\gamma_{max}$ values. This combination, together with an increased particle density during the flare, provides a strong flux increase that is limited mostly to the high-energy bands, consistent with the observed small fluctuations at lower energies.

The parameters for the injection scenario are: $R=5.5 \times 10^{15}$\,cm, $B=0.2$\,G, $\gamma_{min}=5 \times 10^3$, $\gamma_{max}=8 \times 10^5$. The same radius was used for the Fermi-I reacceleration scenario, but with a smaller magnetic field of $B=0.05$\,G and a smaller value of $\gamma_{max}=1.5 \times 10^{6}$ for the initial electron spectrum. Particle acceleration was achieved with an underlying weak turbulent acceleration ($\mathcal{L}=0.4$, $\lambda_{max}=0.1$\,R) and a dominant shock acceleration with timescale $t_{F_I}=1.6$\,R/c.
The best scenario of Fermi-II reacceleration invoked a smaller radius of $R=4.1 \times 10^{15}$\,cm, an intermediate magnetic field strength $B=0.1$\,G, $\gamma_{min}=2 \times 10^3$ and $\gamma_{max}=2 \times 10^6$ for the initial injection. Particle acceleration during the flare was achieved with $\mathcal{L}=0.6$ and $\lambda_{max}=0.1$\,R. 

In all scenarios, the flare duration was adjusted to match the observed duration of the flux variation. For the acceleration scenarios, the approximate reproduction of the LCs is a two-step process and the result depends as much on the choice of the poorly constrained initial parameters of the flaring zone, as on the acceleration timescales that define the flare evolution. 

The best representation of the flare under study is given by the injection scenario,
which is also compatible with a Fermi I acceleration scenario in the limit of very efficient acceleration, as shown in Section \ref{subsec::char2a}. The Fermi II reacceleration scenario does not provide an acceptable shape in the X-ray band for the given constraints on flare amplitude and rise and fall times. The Fermi I reacceleration scenario does not reproduce the rise in the X-ray flare sufficiently well and would require a second
flare component to explain the data points after the observational gap in both bands. 
While we do not claim that this application provides an overall interpretation of the time-dependent multi-band emission, it successfully reproduces the flux variations in the two well sampled high-energy bands and, most importantly, it shows that the different one-zone scenarios present characteristic and distinguishable shapes.

It should be noted that the available data set of well-sampled rapid multiwavelength flares of high-frequency peaked BL Lac objects is very limited. A recent extension of the EMBLEM code to include external photon fields \citep{Dmytriiev2024} will enable future applications to the richer flare data sets from flat-spectrum radio quasars (FSRQs). The inclusion of radiation from the accretion disk, broad line region, or dust torus would introduce additional IC interactions, resulting in a stronger CD in the SED, which would be necessary to reproduce the SEDs of FSRQs and possibly low-frequency peaked BL Lac objects. It would also lead to stronger IC cooling, which limits the maximum electron energies in our scenarios. For constant radiation fields, the different scenarios will keep their temporal characteristics. In the case of a moving blob that could encounter variable radiation from external fields due to its proper motion \citep[e.g.,][]{LeBihan2025}, an additional effect would need to be considered on the evolution of the LC.

\section{Conclusions and outlook}
\label{sec::conclusions}

Even in the simple framework of the single-zone SSC model, different assumptions on the evolution of the particle distribution based on a few generic injection and acceleration scenarios can lead to a large variety of flare shapes. The injection, Fermi-I and Fermi-II (re)acceleration scenarios explored in this study are characterized by distinct observational signatures across different energy bands, which can be tested with multiwavelength data, such as the times of flare peaks and onsets, rise times, the existence of a plateau during the flare or the signature of a hysteresis in the SED.

The Fermi-I and Fermi-II (re)acceleration scenarios show the most visible changes in LC shapes and delays between bands. In particular, the onset and peak times of flares vary between scenarios, with the acceleration scenarios showing the largest delays between low and high energy bands. The reacceleration scenarios moreover exhibit the strongest hysteresis effects, especially in the synchrotron peak. To try to discriminate between the scenarios requires data in at least two energy bands. The complexity of direct observational comparison with these models is especially notable in the Fermi-II scenarios, which are particularly sensitive to parameter changes, with small adjustments leading to significant differences in flare profiles.

An illustrative application to a 2013 flare of Mrk\,421 shows both the potential and the challenges of applying such models to real data. 
While this application does not reproduce the full time-dependent multi-band behavior of the source, each scenario leads to distinguishable shapes in the high-energy bands that can provide specific insight into the temporal and spectral behavior of the flare.
For the given data set, the high-energy flare shape seems to be best reproduced by the simple injection scenario, which is indistinguishable from our simplified treatment of the shock-acceleration scenario for the most efficient acceleration.
 
The identification of generic features expected in the light curves of rapid blazar flares will be of particular interest for future observations with the Cherenkov Telescope Array Observatory (CTAO). Its proposed Target of Opportunity programme \citep{Samarai2019} will benefit from its largely increased flux sensitivity and thus temporal resolution compared to current Imaging Air Cherenkov Telescope arrays. Since CTAO will cover the VHE band from a few 10\,GeV to a few 100\,TeV, it will record densely sampled light curves of the emission from the most energetic particle population, accessing the most rapid cooling timescales. It will thus be particularly sensitive to flux and spectral variations expected from the microscopic processes modeled in this work. These observations should be complemented by MWL data through dedicated campaigns to be able to distinguish different scenarios.

Future work should incorporate external photon fields, for applications to the broader data sets of FSRQ flares. The obvious limitations of the one-zone model could be overcome by implementing a series of acceleration and emission zones permitting one to better approximate the diffusion of particles from the locations of the accelerators. Multi-zone models exist for specific scenarios \citep[e.g.,][]{Marscher2014,Zhang2016}, but do not allow a systematic comparison of different acceleration processes to the best of our knowledge. 

Another effect that should be evaluated for a more complete physical treatment of scenarios 3a and 3b is the damping of turbulence induced by the electrons that are accelerated through resonant interactions. \citet{Tavecchio2022} show that for the steady-state emission of extreme BL Lac objects, the turbulence damping time is shorter than the cascading time of the turbulence, so that the interplay between particles and turbulence should be modeled. Damping is expected to lead to a softer SSC spectrum and lower Compton dominance. 
We leave this effect to a future dedicated study, where a more self-consistent treatment based on energy conservation, as applied by \citet{Sciaccaluga2022} or, in a more general description, by \citet{Lemoine2024} will allow us to estimate its impact on the Fermi II (re)acceleration LCs.
Following \citet{Kakuwa2015}, we note however that damping may be negligible if stochastic acceleration is not driven by Alfvén waves, but by fast mode waves.

One scenario not considered here, which is promising in particular in view of the most rapid flares observed at the minute timescale in the VHE band, is acceleration through magnetic reconnection \citep[e.g.,][]{Giannios2009,Gouveia2010}. This scenario would require a very different environment, i.e.,\ a magnetically dominated emission region, probably located close to the jet base or in locally magnetized regions along the jet, which is generally difficult to reconcile with SSC modeling of SEDs from BL Lac objects. 
In certain regimes, especially when it is mediated by turbulence, the process becomes independent of resistivity and leads to multiple simultaneous reconnection “flashes,” which can phenomenologically be described as a first-order process similar to shock acceleration \citep[e.g.,][]{Kowal2012}. This process could thus be modeled approximately as an injection of a power-law or kappa particle distribution \citep[cf.][]{Aimar2023}.

Given the limited sample of well-observed multiwavelength BL Lac flares, expanding both the observational scope and theoretical modeling will be crucial to disentangle between the processes at play in the large variety of blazar flaring events.

\begin{acknowledgements}
       This work has received support under the programme «Investissements d’Avenir » launched by the French Government and implemented by ANR with the reference ANR-20-SFRI-0010 PSL. AD acknowledges support from the Department of Science and Innovation and the National Research Foundation of South Africa through the South African Gamma-Ray Astronomy Programme (SA-GAMMA).\\ \textit{Data availability.} The additional figures are publicly available on Zenodo at \url{https://doi.org/10.5281/zenodo.18430257}.
\end{acknowledgements}

\begin{appendix}

\section{Numerical scheme of the EMBLEM code}
\label{app:numerical}
We describe the numerical scheme used to solve the kinetic equation with the EMBLEM code, as described by \cite{Dmytriiev2021} and including the changes brought to simulate adiabatic expansion and time- and energy-dependent Fermi II acceleration. The code solves equation \ref{Fok_P_eq1} using a fully implicit difference scheme detailed in \cite{Chang1970}, by expressing it using the coefficients $V1$, $V2$, $V3$ and the terms $S$:
\begin{equation}
    V1_j n_{j-1}^{k+1} + V2_j n_j^{k+1} + V3_j n_{j+1}^{k+1} = S_j^k
\end{equation}
where $n(x,t)$ is the particle distribution function, satisfying $N_e=\int_0^{\infty} n(x,t)dx$. The momentum $x$ is defined on a non-equidistant grid $\{x_j\}$ while the time grid is homogeneous so that $t_k=k\Delta t$ with $\Delta t=(t_{end}-t_{start})/(\mathcal{K}-1$). We define $t_{start}$ and $t_{end}$ as the beginning and ending times of the simulation, while $\mathcal{K}$ is the number of points on the time grid. The particle distribution function is thus sampled with the set $n_j^k=n(x_j,t_k)$. The Lorentz factor grid, as prescribed by \cite{Chiaberge1999}, is defined with equidistant logarithmic steps:
\begin{equation}
    \gamma_j=\gamma_{min}\left(\frac{\gamma_{max}}{\gamma_{min}}\right)^{(j-1)/(J-1)}
\end{equation}
with $\gamma_{min}$ and $\gamma_{max}$ the minimal and maximal Lorentz factor for computation, respectively, and $J$ the number of points on the Lorentz factor grid.
\\The left-hand side coefficients of the kinetic equation are:
\begin{align}
    V1_j &= -\frac{\Delta t}{\Delta \gamma_j}\frac{C_{j-1/2}}{\Delta\gamma_{j-1/2}}W_{j-1/2}^-\\
    V2_j &=1 + \Delta t\left(\frac{1}{t_{esc}}+\frac{3}{t_{ad}}\right)+\frac{\Delta t}{\Delta\gamma_j}\left(\frac{C_{j-1/2}}{\Delta\gamma_{j-1/2}}W_{j-1/2}^+ + \frac{C_{j+1/2}}{\Delta\gamma_{j+1/2}}W_{j+1/2}^-\right)\\
    V3_j &= -\frac{\Delta t}{\Delta\gamma_{j}}\frac{C_{j+1/2}}{\Delta_{j+1/2}}W_{j+1/2}^+
\end{align}
where we have introduced the notation $n_{j+1/2}^{k+1}=(1-\delta_j)n_{j+1}^{k+1}+\delta_jn_j^{k+1}$ and the following weights: 
\begin{align}
    W_{j\pm1/2}^\pm &= \frac{w_{j\pm1/2}\exp(\pm w_{j\pm1/2}/2)}{2\sinh
    (w_{j\pm 1/2}/2)}\\
    w_{j\pm1/2} &= \frac{B_{j\pm 1/2}}{C_{j\pm 1/2}}\Delta\gamma_{j\pm1/2}\\
    \Delta\gamma_{j\pm 1/2} &= \gamma_{j\pm1/2+1/2} - \gamma_{j\pm1/2-1/2}\\
\end{align}
Here, $B$ and $C$ are the coefficients containing the timescales for acceleration and cooling:
\begin{align}
    B(\gamma,t) &= b_c(\gamma,t)\gamma^2 - \left[a(t)-\frac{1}{t_{ad}}+\frac{2}{t_{F_{II}}}\right]\gamma\\
    C(\gamma,t) &= \frac{(\gamma m_ev)^2}{t_{F_{II}}}
\end{align}
The r.h.s. term is given by $S_j^k=n_j^k+Q_j^k\Delta t$ with $Q_j^k$ the injection term corresponding to a number of particles per unit volume, per unit time and per unit of Lorentz factor interval. We normalize the injection term so that the number of injected electrons remains equal, during the expansion, to the one at time $t=t_{exp}$.

\section{Approximate treatment of particle acceleration on multiple shocks}
\label{app:multishock}

\begin{figure} [h!]
    \centering
    \includegraphics[trim={12mm, 0mm, 12mm, 20mm},clip=true,width=0.9\columnwidth]
    {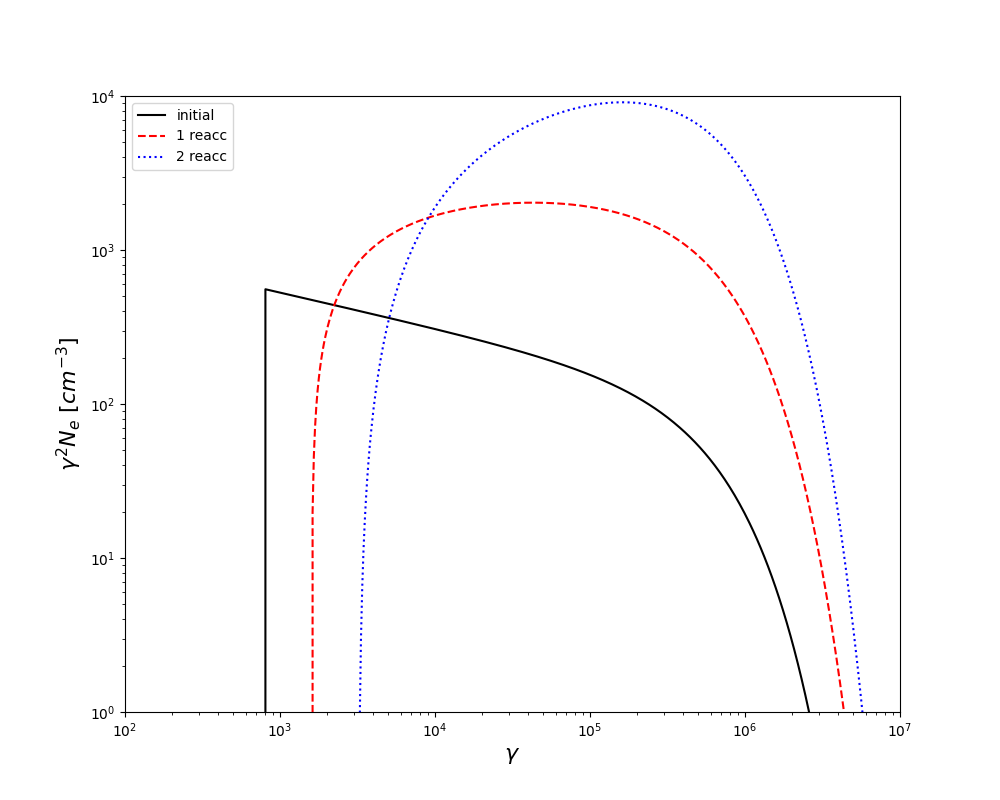}
    \caption{The expected stationary differential electron energy spectrum, multiplied by $\gamma^2$, is shown as function of $\gamma$ for acceleration on a single mildly relativistic shock (black line), on a second (red dashed line), and third idential shock (blue dotted line).}
    \label{appfig:electrons_theory}
\end{figure}

Following \citet{Zech2021} and references therein, the crossing of multiple mildly relativistic shocks by a particle population will lead to an increasing minimum energy and to a hardening spectrum. The authors focused on the simplified case of reacceleration of the whole initial particle population with negligible losses.
Applying their analytical description to the parameters chosen for our models leads to the expected stationary particle distributions shown in Fig.~\ref{appfig:electrons_theory}. 

\begin{figure} [h!]
    \centering
    \includegraphics[trim={12mm, 0mm, 12mm, 20mm},clip=true,width=0.9\columnwidth]
    {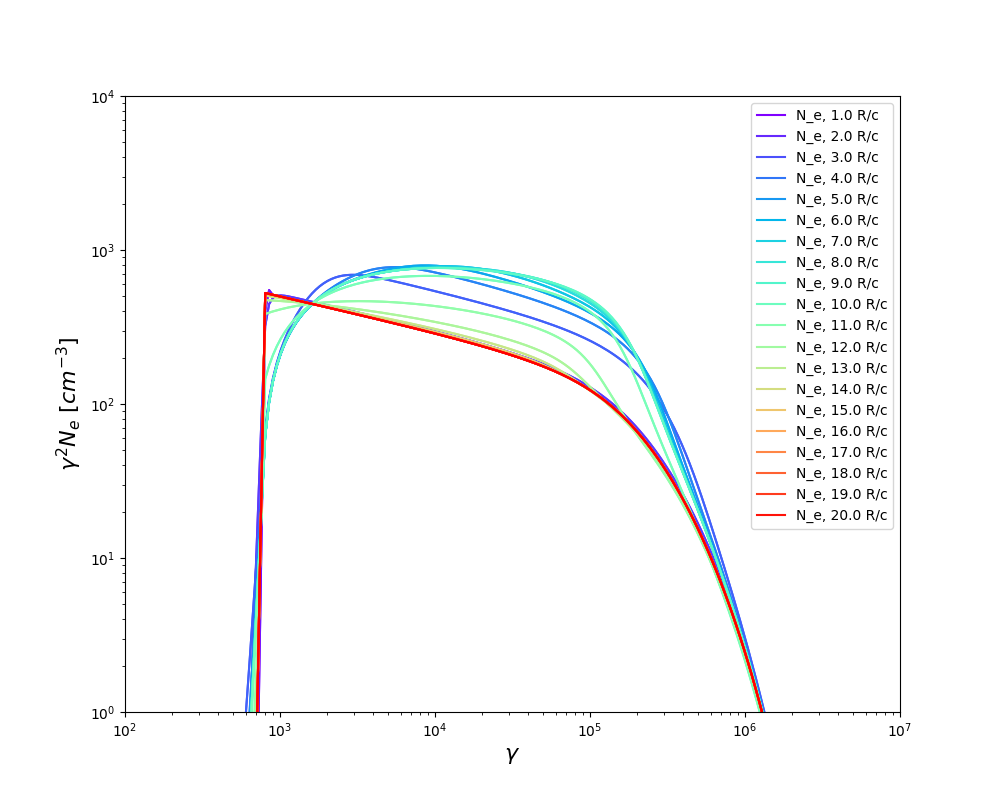}
    \caption{The modeled time-dependent differential electron spectrum, multiplied by $\gamma^2$, shown as function of $\gamma$ for the case of $t_{FI} =2 R/c$.}
    \label{appfig:electrons_model_2}
\end{figure}

To reproduce the two main spectral characteristics of multiple shock acceleration (scenario 2b), we propose a simple empirical approach making use of the term $a(t)$ of systematic energy gain in Equ.~\ref{Fok_P_eq1}. As an example, Fig.~\ref{appfig:electrons_model_2}
shows the temporal evolution of the differential particle spectrum, including escape and cooling,
for an acceleration timescale $t_{FI}=2$\,R/c. The acceleration term is only switched on during the flare duration. Within the parameter region explored in the present paper, this continuous approach results in the wanted spectral behavior.

\section{Input parameters}
\label{app:parameters}

\begin{table}[!h]
\caption[]{Input parameters used in the simulations for the different flare scenarios.}
\centering
\begin{tabular}{ |p{5.5cm}||p{2cm}|p{2cm}|p{2cm}|  }
     \hline
     Parameter& Variable & Value & Unit\\
     \hline
     \multicolumn{4}{|c|}{Source} \\
     \hline
     Initial magnetic field & $B_0$ & 0.04 & G\\
     Magnetic field configuration & $m_{B}$ & 1 & -\\
     Initial blob radius & $R_0$ & $2.8\times 10^{16}$ & cm\\
     Blob Doppler factor & $\delta$ & 29 & -\\
     Redshift & $z$ & 0.0308 & -\\
     Angle to observer & $\theta$ & \{0 ; 1\} & $\degree$\\
     \hline
     \multicolumn{4}{|c|}{Evolution} \\
     \hline
     Quiescent injection duration & $t_{\rm inj}$ & 30 &  R/c\\
     Quiescent escape timescale & $t_{\rm esc}$ & 1 & R/c\\
     Beginning of expansion & $t_{\rm exp}$ & 0 & R/c\\
     Beginning of flaring & $t_{\rm flare}$ & 2 & R/c\\
     Duration of flaring & $t_{\rm dur}$ & 3 & days*\\
     Escape timescale during flaring & $t_{\rm esc,flare}$ & 1 & R/c\\
     \hline
     \multicolumn{4}{|c|}{Continuous injection spectrum} \\
     \hline
     Spectrum normalization & $\Dot{N}_{\rm inj}$ & $1.86\times 10^{-14}$ & cm$^{-3}$s$^{-1}$\\
     Spectrum slope & $\alpha_{\rm inj}$ & -2.23 & -\\
     Pivot Lorentz factor & $\gamma_{\rm inj,pivot}$ & $1.0 \times 10^5$ & -\\
     Cutoff Lorentz factor & $\gamma_{\rm inj,cut}$ & $5.8 \times 10^5$ & -\\
     Minimal injected Lorentz factor & $\gamma_{\rm inj,min}$ & 800 & -\\
     \hline
     \multicolumn{4}{|c|}{Flaring injection spectrum} \\
     \hline
     Spectrum normalization & $\Dot{N}_{\rm add}$ & $4\times 10^{-14}$ & cm$^{-3}$s$^{-1}$\\
     Spectrum slope & $\alpha_{\rm add}$ & -2.23 & -\\
     Pivot Lorentz factor & $\gamma_{\rm add,pivot}$ & $1.0 \times 10^5$ & -\\
     Cutoff Lorentz factor & $\gamma_{\rm add,cut}$ & $5.8 \times 10^5$ & -\\
     Minimal injected Lorentz factor & $\gamma_{\rm add,min}$ & 800 & -\\
     \hline
     \multicolumn{4}{|c|}{Particle acceleration} \\
     \hline
     Fermi I reacceleration timescale & $t_{\rm F_I}$ & $[0.7-2.0]$ & R/c\\
     Turbulent spectrum slope & $q$ & 2 & -\\
     Maximum spectrum wavelength  & $\lambda_{\rm max}$ & $[0-1]$ & R\\
     Turbulence level & $\mathcal{L}_{\rm turb}$ & $[0-1]$ & -\\
     \hline
     \multicolumn{4}{|c|}{Simulation grids} \\
     \hline
     Minimal electron Lorentz factor & $\gamma_{\rm min}$ & 100 & -\\
     Maximal electron Lorentz factor & $\gamma_{\rm max}$ & $1.0\times 10^{7}$ & -\\
     Lorentz factor grid's points & $n_{\rm \gamma}$ & 200 & -\\
     Simulation start time & $t_{\rm start}$ & 0 & R/c\\
     Simulation end time & $t_{\rm end}$ & 30 & R/c\\
     Time grid's bins & $n_{\rm bins}$ & 1000 & -\\
     \hline
\end{tabular}

\vspace{0.3cm}
\tablefoot{Curly brackets indicate different choices made for a parameter depending on the scenario. *The duration of flaring is given in units of days in the observer frame.}
\label{table_parameters}
\end{table}

\clearpage

\section{Spectral energy distributions}
\label{app:seds}

\begin{figure}[!h]
    \vspace{-0.5cm}
    \centering
    \includegraphics[trim={0.5cm 0.cm 0.5cm 1cm},clip,width=0.85\columnwidth]
    {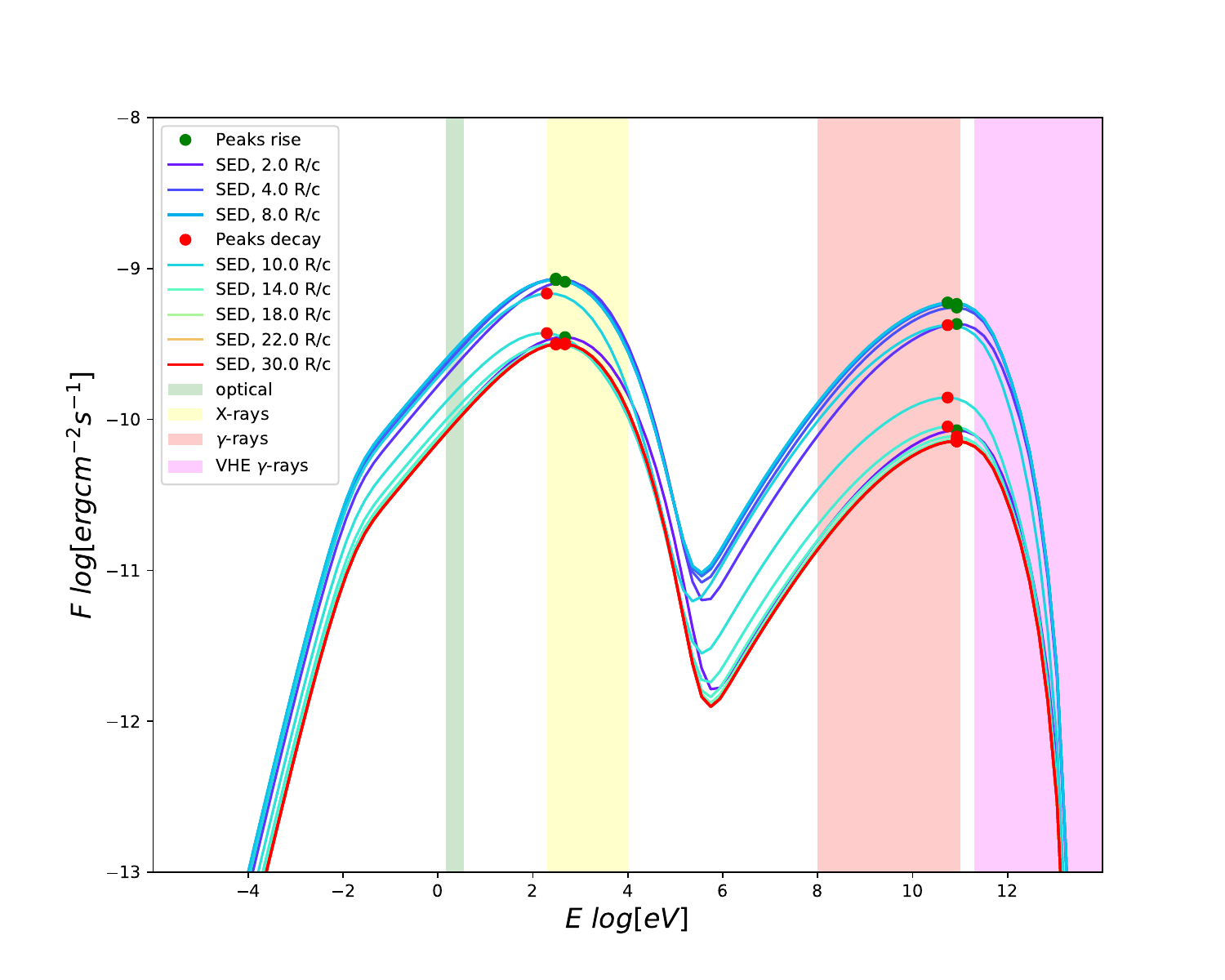}
    \caption{Time-evolving SED of scenario 1a, for $\dot{N}_{add}=4\times 10^{-14}$ cm$^{-3}$s$^{-1}$. The colored curves represent the temporal evolution. The vertical bands correspond to the area integrated to compute the LC in four energy bands.
    }
    \label{appfig:sed_scenario1a}
\end{figure}

\begin{figure} [!h]
    \vspace{-0.5cm}
    \centering
    \includegraphics[trim={0.5cm 0.cm 0.5cm 1cm},clip,width=0.85\columnwidth]
    {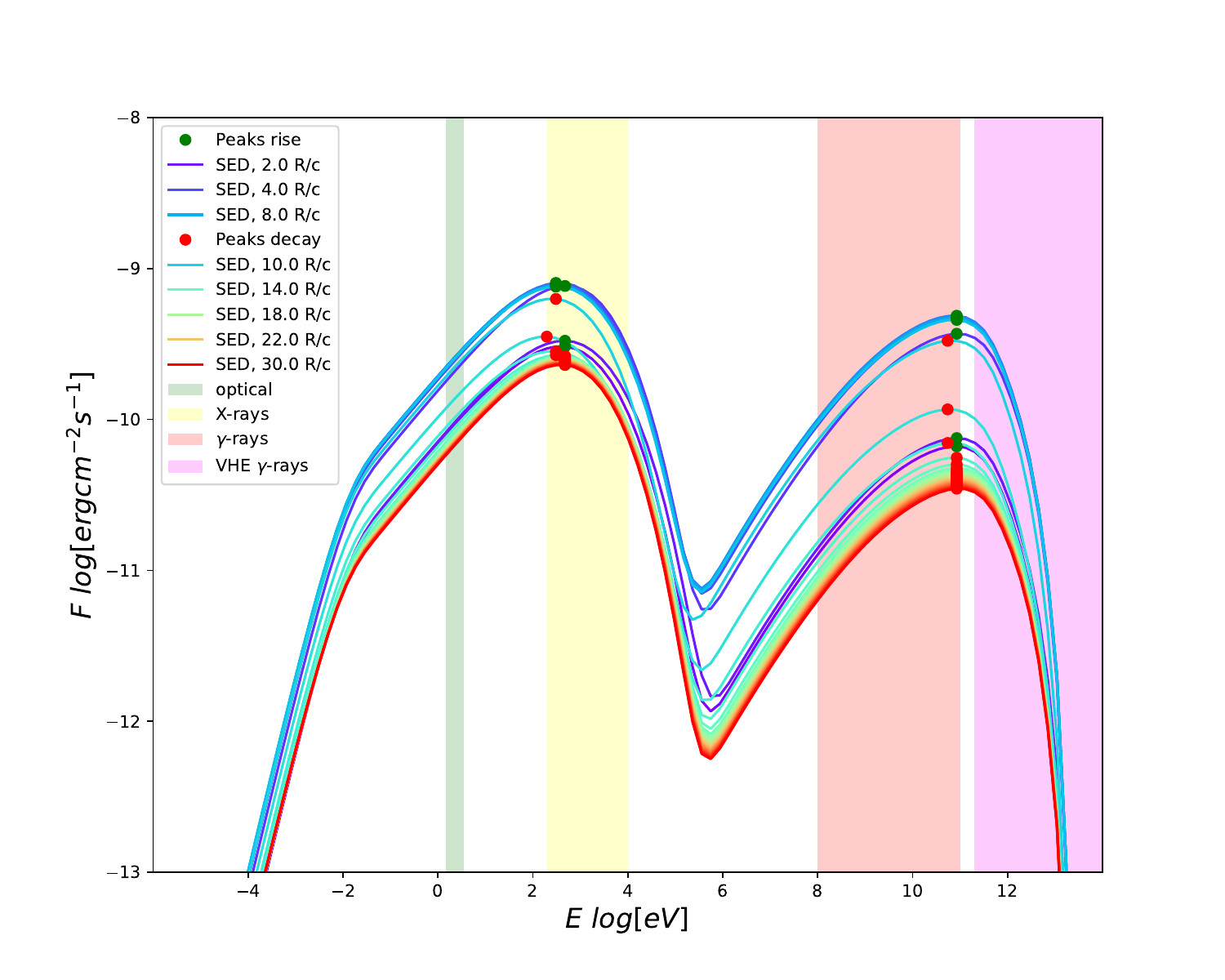}
    \caption{Time-evolving SED of scenario 1b, for $\dot{N}_{add}=4\times 10^{-14}$ cm$^{-3}$s$^{-1}$.}
    \label{appfig:sed_scenario1b}
\end{figure}

\begin{figure} [!h]
   \vspace{-0.5cm}
    \centering
    \includegraphics[trim={0.5cm 0.cm 0.5cm 1cm},clip,width=0.85\columnwidth]
    {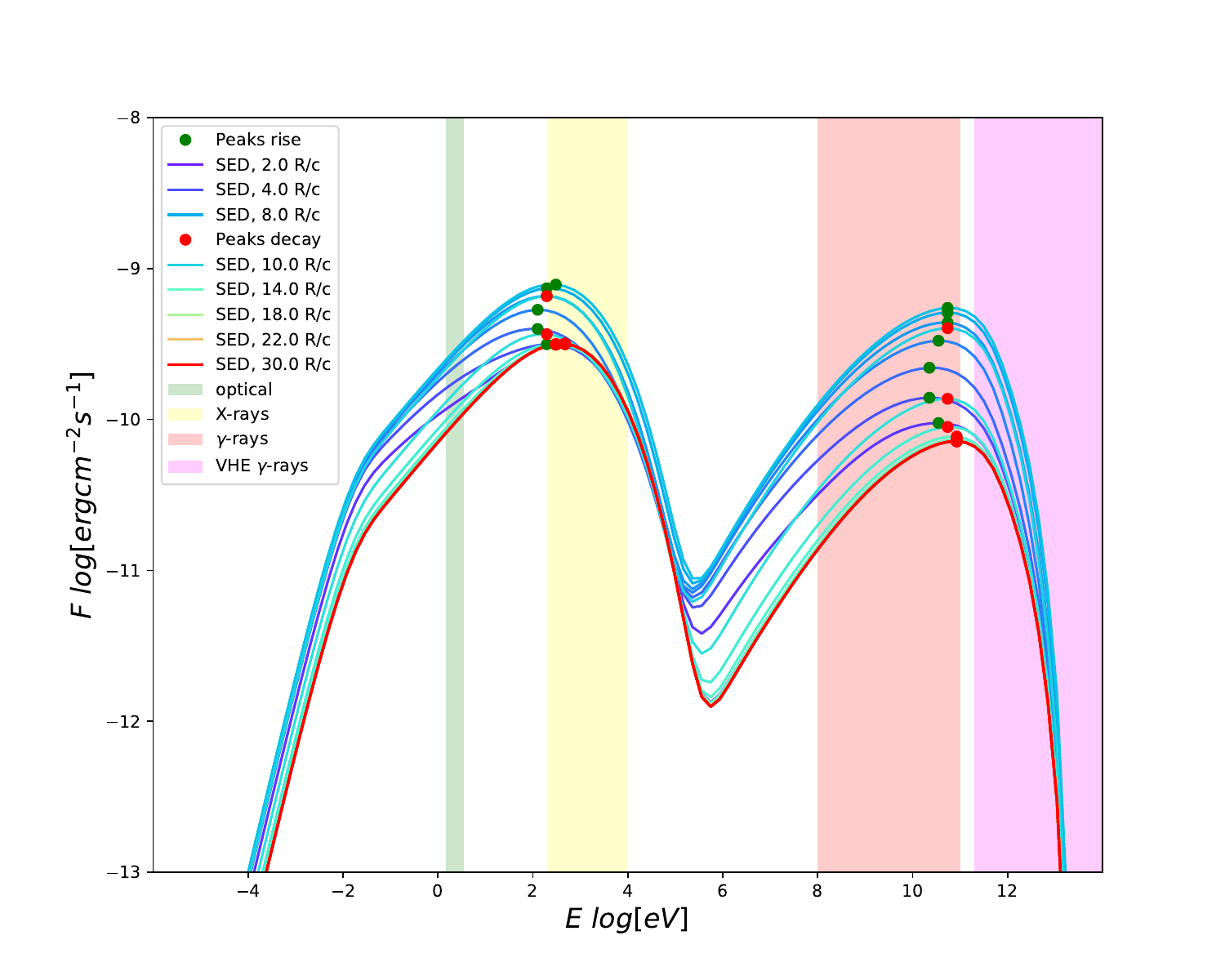}
    \caption{Time-evolving SED of scenario 2a, for $t_{\rm shock} = 1.0$ R/c. }
    \label{appfig:sed_scenario2a}
\end{figure}

\begin{figure} 
    \centering
    \includegraphics[trim={0.5cm 0.cm 0.5cm 1cm},clip,width=0.85\columnwidth]
    {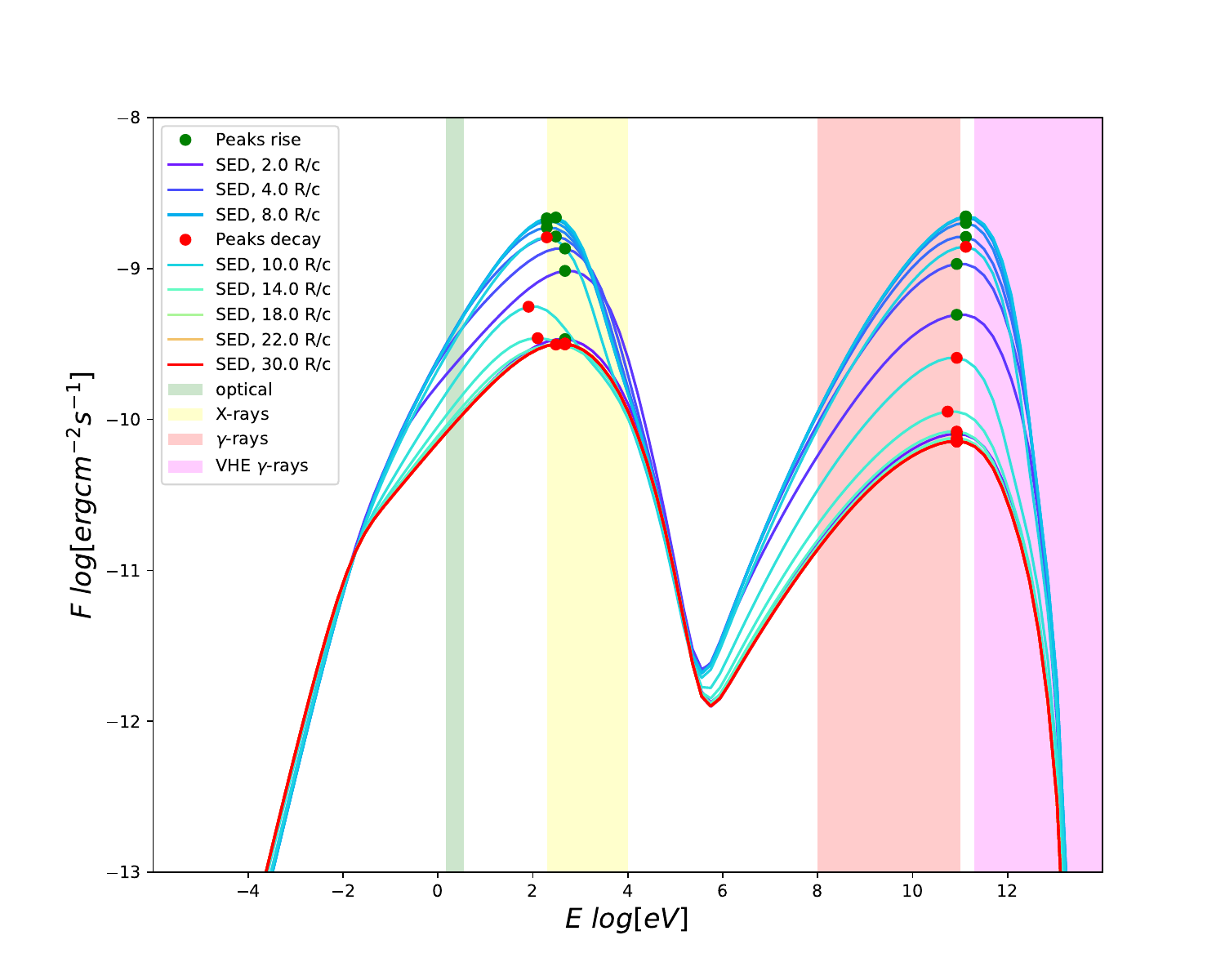}
    \caption{Time-evolving SED of scenario 2b, for $t_{\rm F_I} =1.3$ R/c.}
    \label{appfig:sed_scenario2b}
\end{figure}

\begin{figure} 
    \centering
    \includegraphics[trim={0.5cm 0.cm 0.5cm 1cm},clip,width=0.85\columnwidth]
    {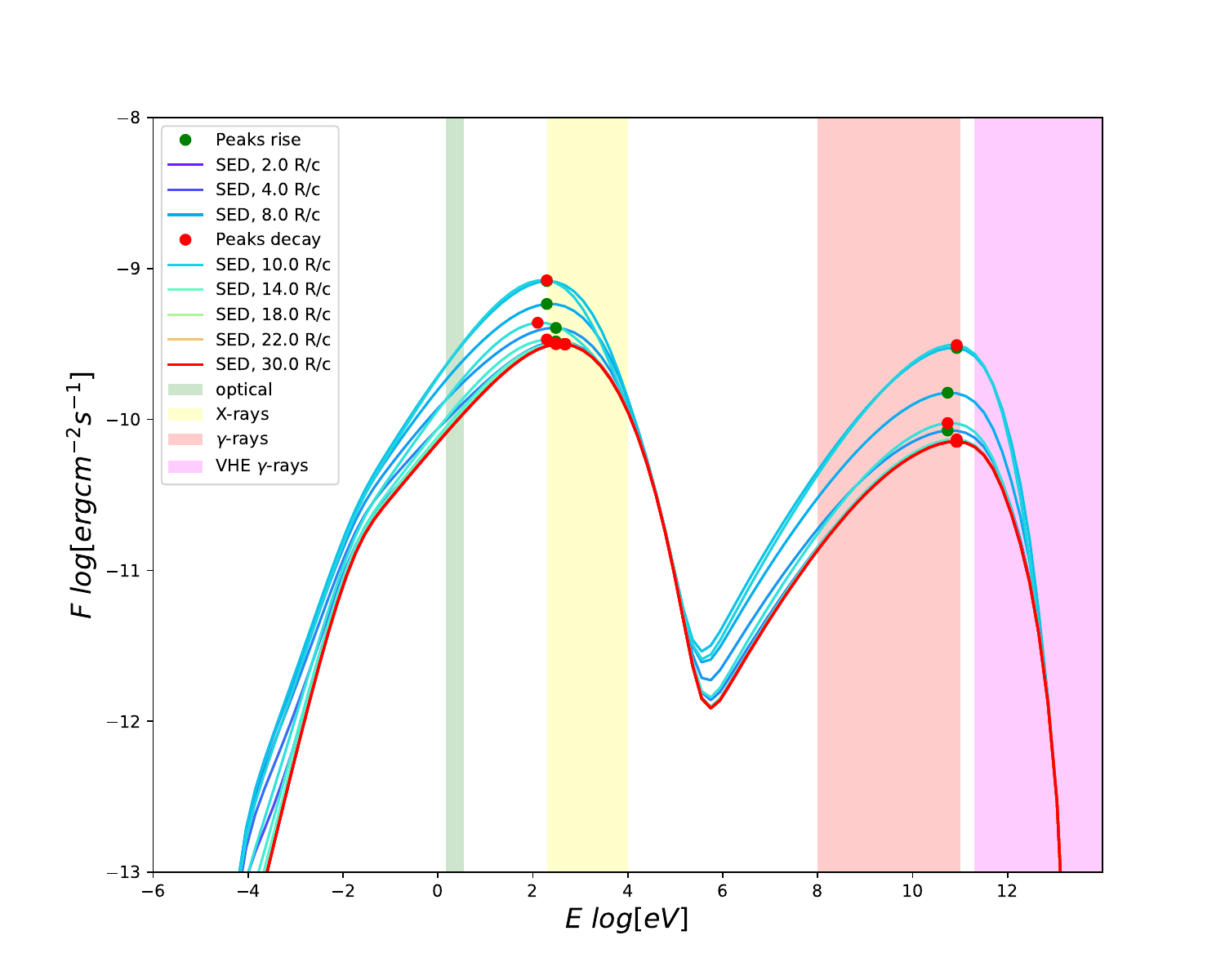}
    \caption{Time-evolving SED of scenario 3a, for $t_{\rm FII} = 3.23 \, \rm{R/c}$.}
    \label{appfig:sed_scenario3a}
\end{figure}

\begin{figure} 
    \centering
    \includegraphics[trim={0.5cm 0.cm 0.5cm 1cm},clip,width=0.85\columnwidth]
    {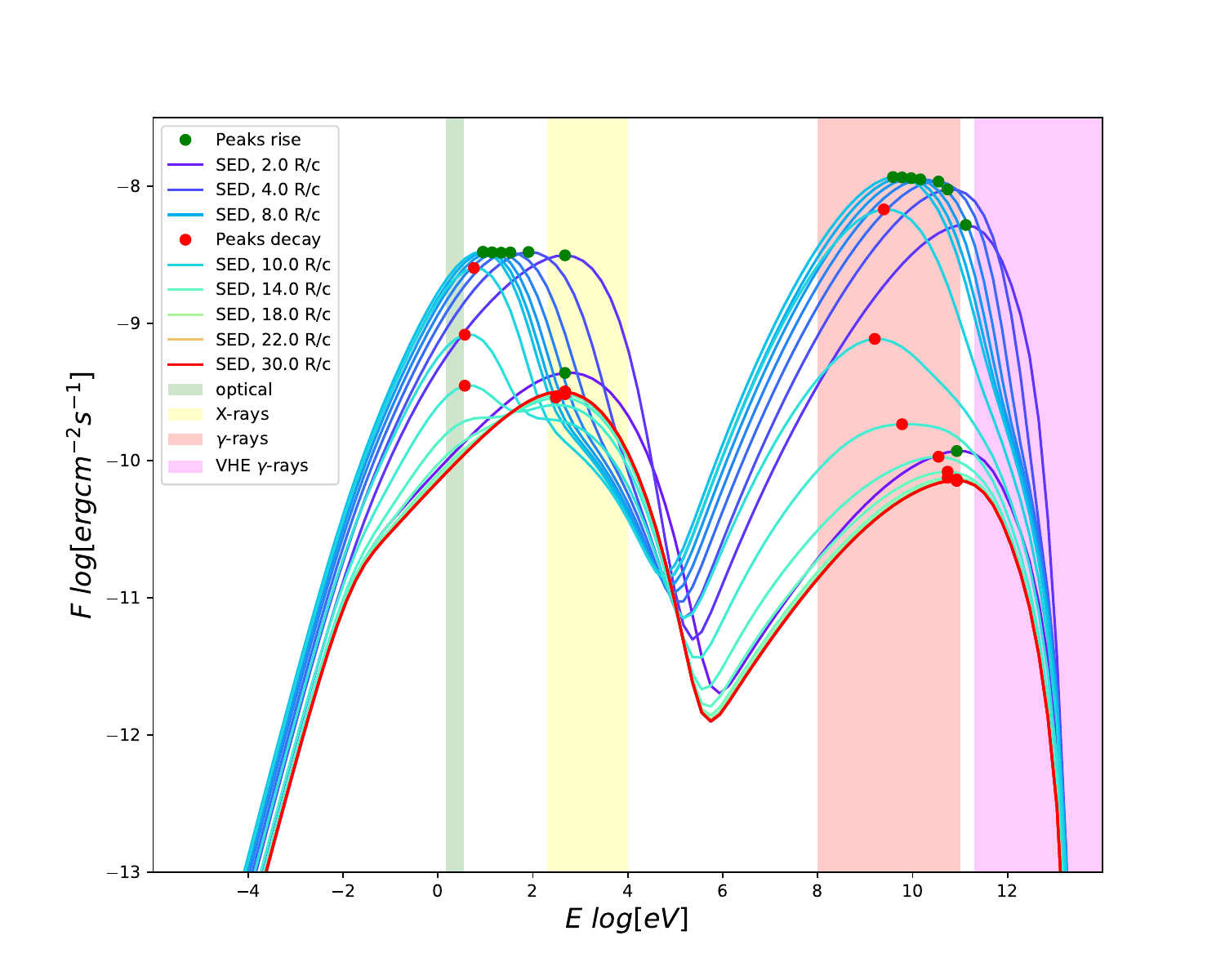}
    \caption{Time evolving SED of scenario 3b, for $t_{F_{II}} = 1.4$ R/c.}
    \label{appfig:sed_scenario3b}
\end{figure}

\FloatBarrier
\clearpage

\section{Light curves in different energy bands}
\label{app:lcs}

\begin{figure}[!h]
    \centering
    \vspace{-0.6cm}
    \includegraphics[trim={0.5 21 0.5 0.5cm},clip,width=0.8\columnwidth]{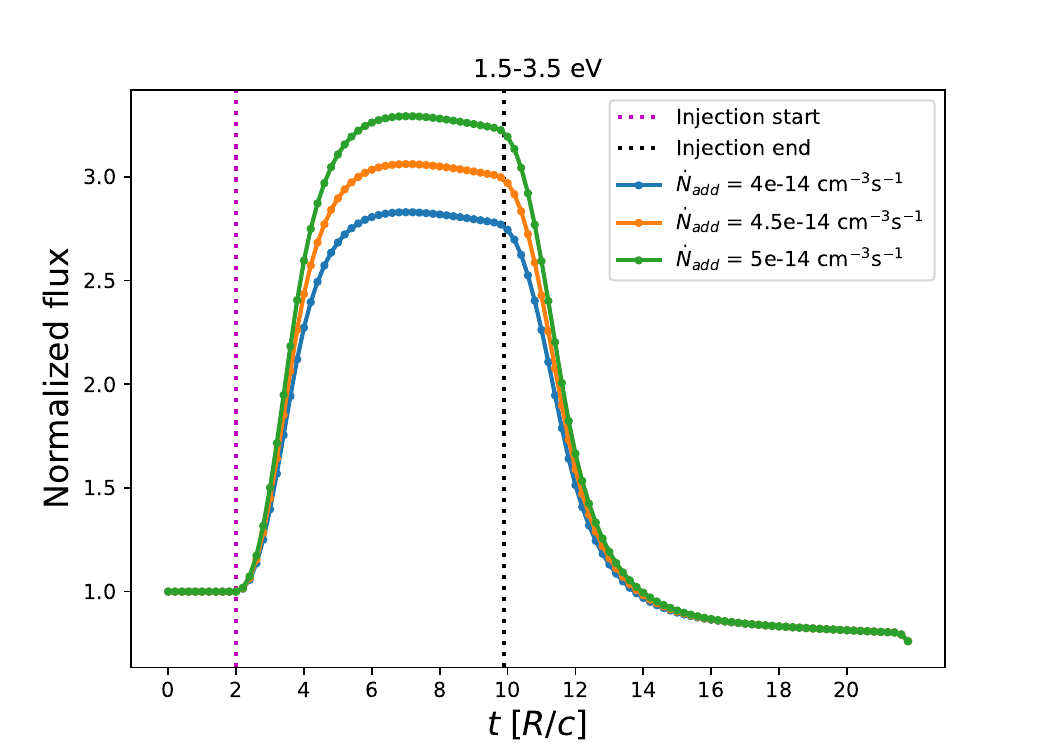}
    \includegraphics[trim={0.5 1.5 0.5 0.5cm},clip,width=0.8\columnwidth]{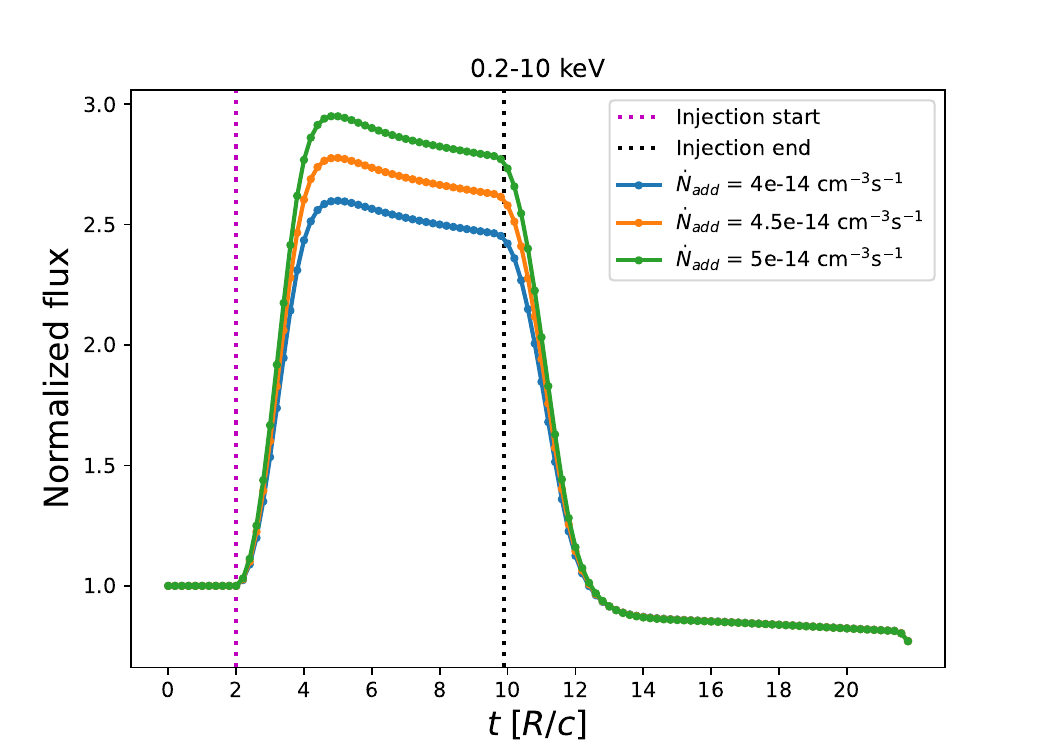}
    \caption{Optical and X-ray LC of flares resulting from additional particle injection with different rates with adiabatic expansion (scenario 1b).}
    \label{appfig:LCs_inject_adiab_compare}
\end{figure}

\begin{figure}
    \centering
    \vspace{-0.3cm}
    \includegraphics[trim={0.5 21 0.5 0.5cm},clip,width=0.8\columnwidth]{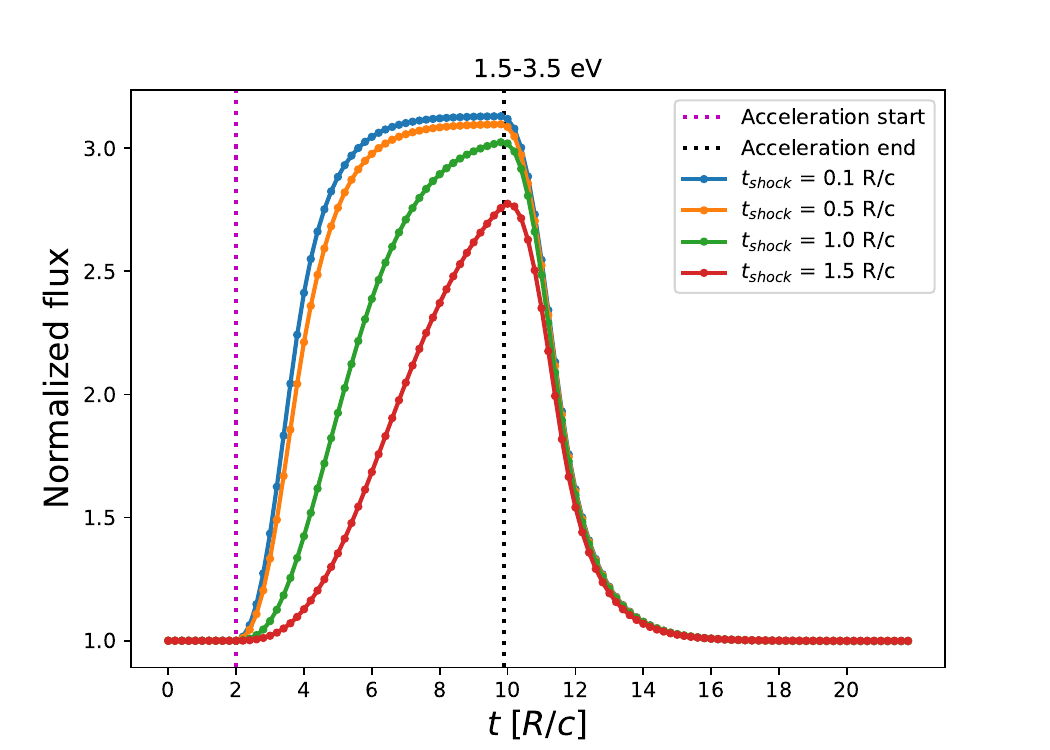}
    \includegraphics[trim={0.5 1.5 0.5 0.5cm},clip,width=0.8\columnwidth]{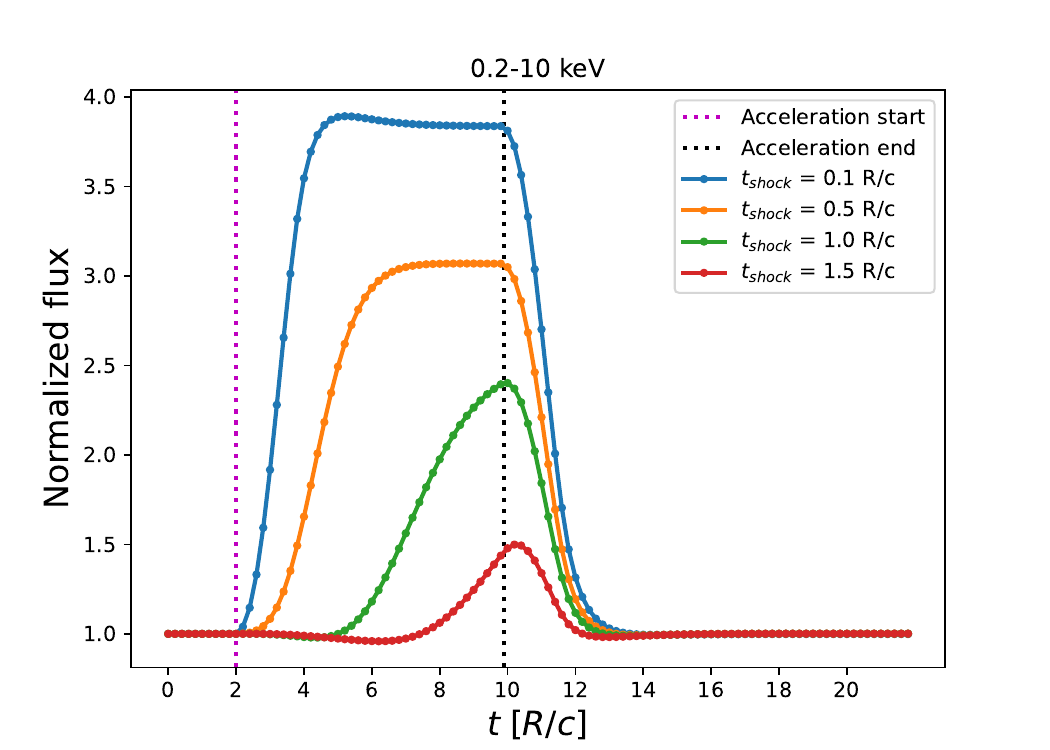}
    \caption{Optical and X-ray LC of flares resulting from DSA for different shock timescales (scenario 2a).}
    \label{appfig:LCs_FI_acc_compare}
\end{figure}

\begin{figure}
    \centering
    \includegraphics[trim={0.5 21 0.5 0.5cm},clip,width=0.8\columnwidth]{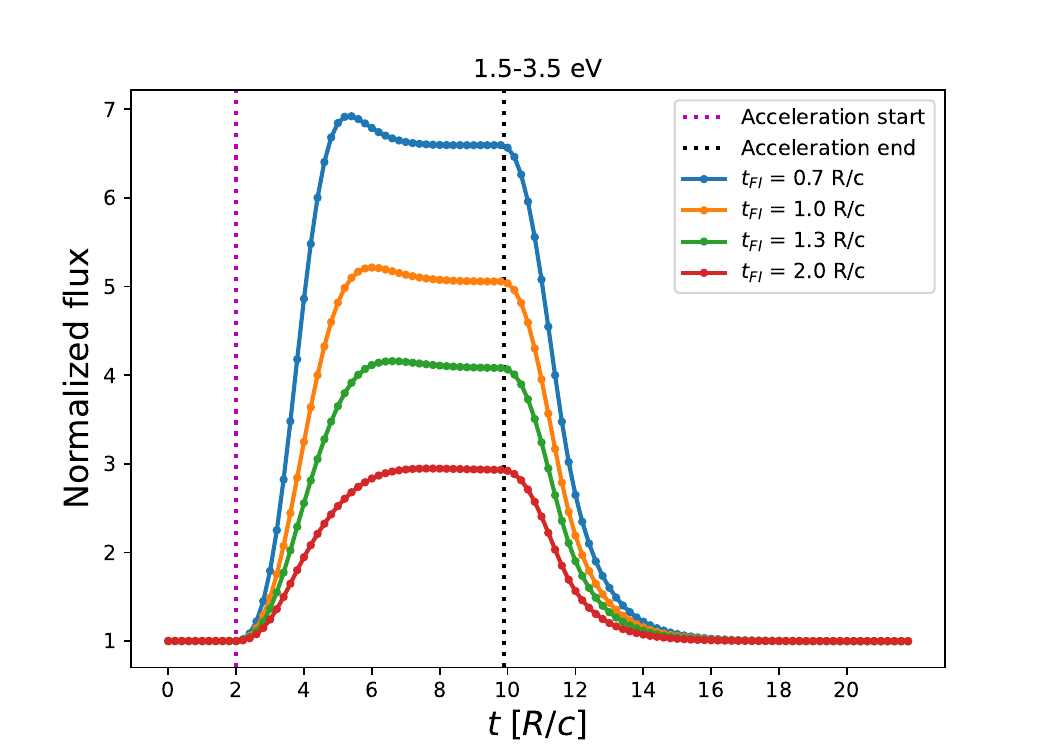}
    \includegraphics[trim={0.5 1.5 0.5 0.5cm},clip,width=0.8\columnwidth]{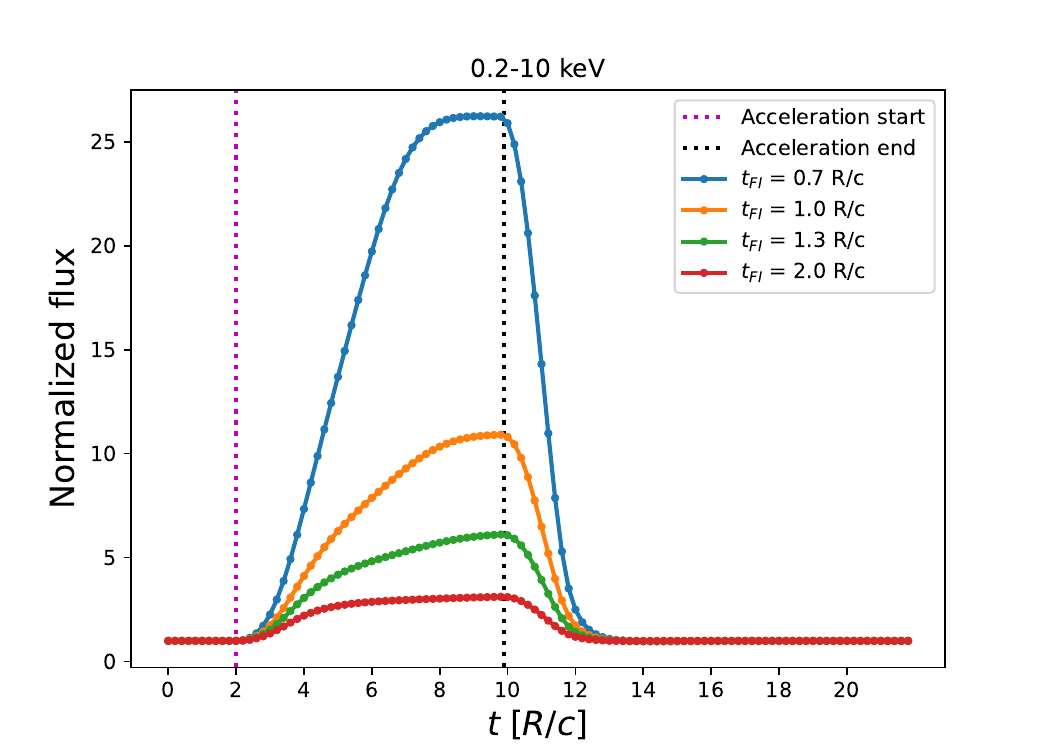}
    \caption{Optical and X-ray LC of flares resulting from Fermi I reacceleration for different timescales $t_{F_I}$ (scenario 2b).}
    \label{appfig:LCs_FI_reacc_compare}
\end{figure}

\begin{figure}
    \centering
    \vspace{-0.02cm}
    \includegraphics[trim={0.5 21 0.5 0.5cm},clip,width=0.8\columnwidth]{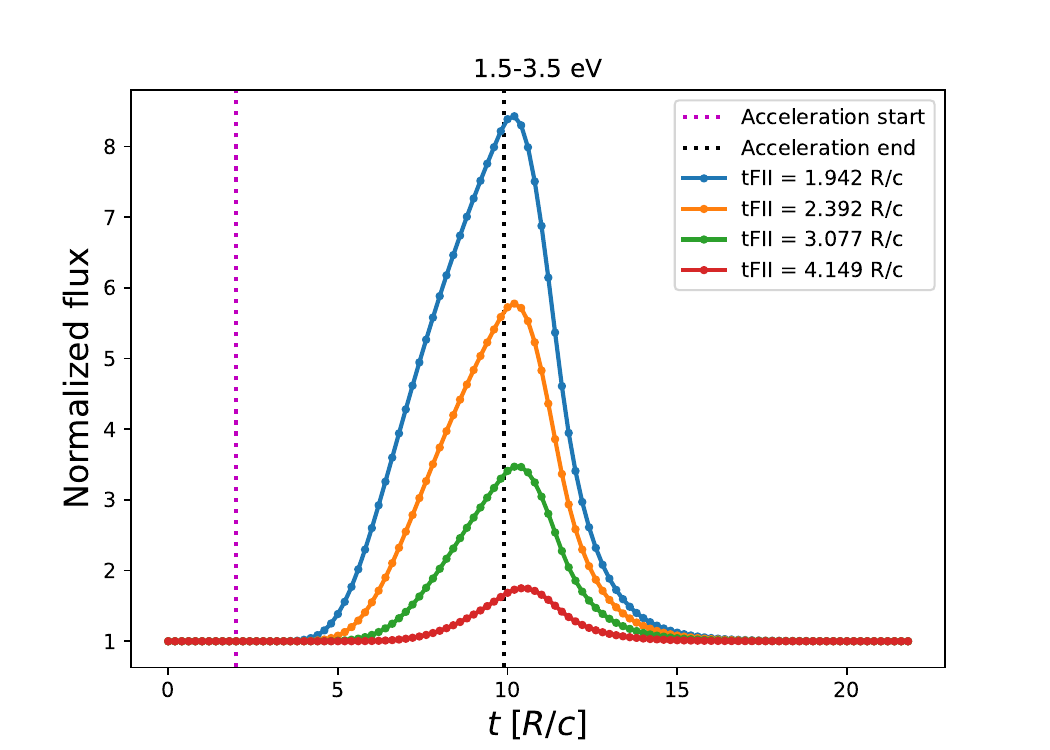}
    \includegraphics[trim={0.5 1.5 0.5 0.5cm},clip,width=0.8\columnwidth]{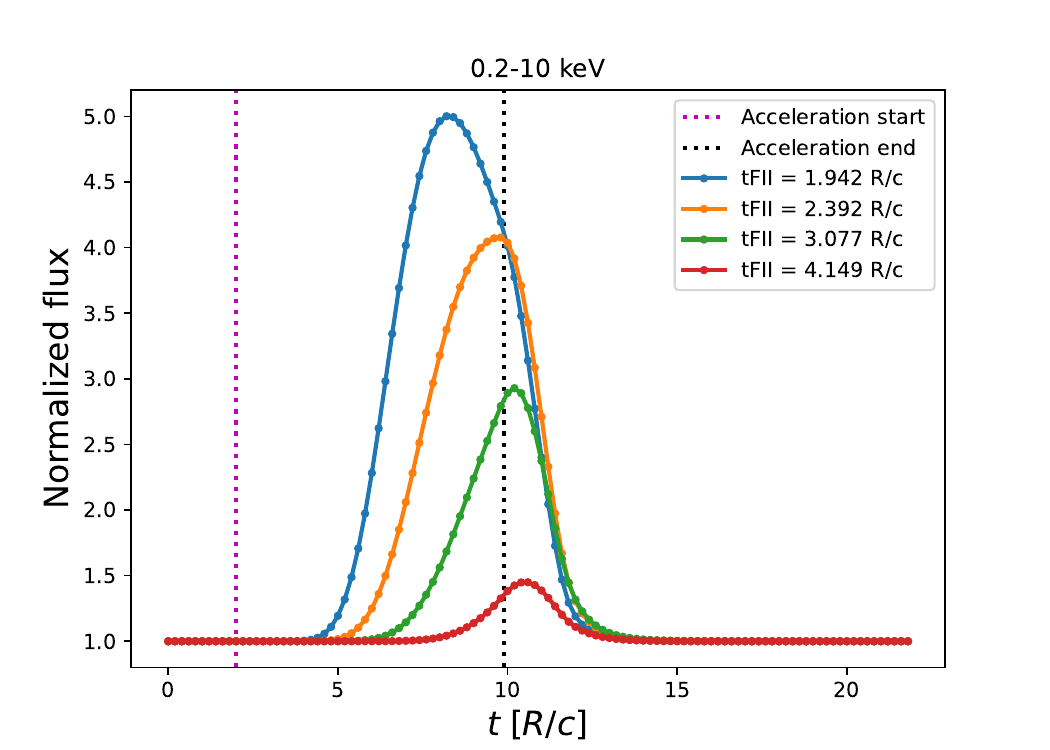}
    \caption{Optical and X-ray LC of flares resulting from hard-sphere Fermi II acceleration for different acceleration timescales $t_{F_{II}}$ (scenario 3a).}
    \label{appfig:LCs_FII_acc_compare}
\end{figure}

\begin{figure}
    \centering
    \includegraphics[trim={0.5 21 0.5 0.5cm},clip,width=0.8\columnwidth]{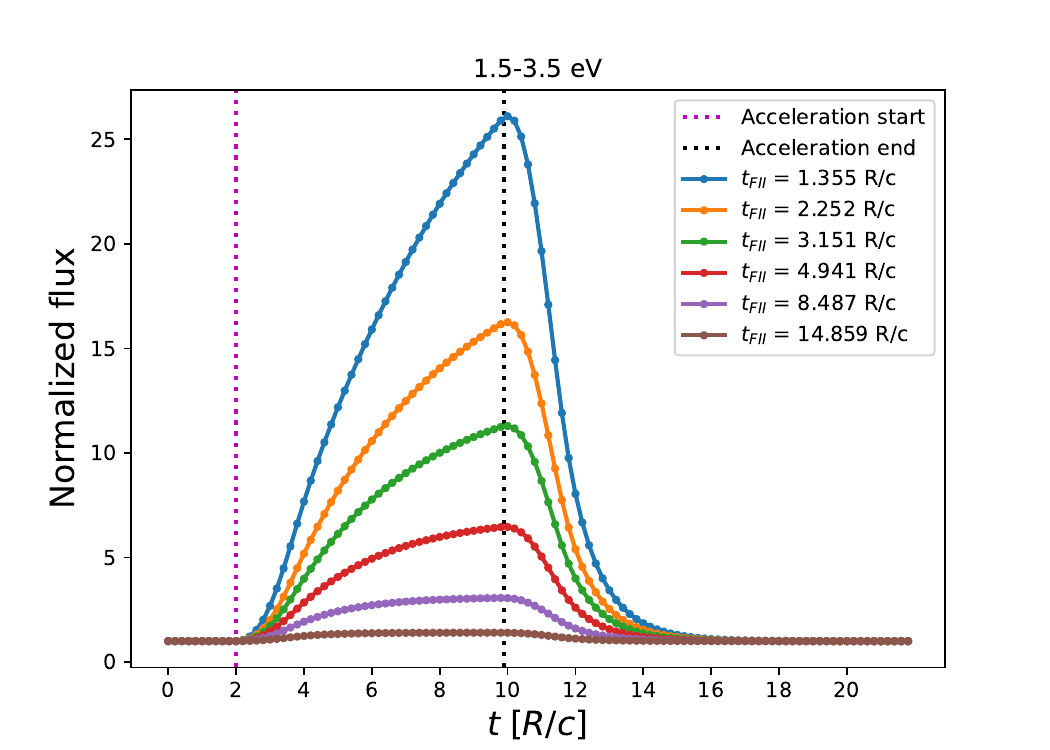}
    \includegraphics[trim={0.5 1.5 0.5 0.5cm},clip,width=0.8\columnwidth]{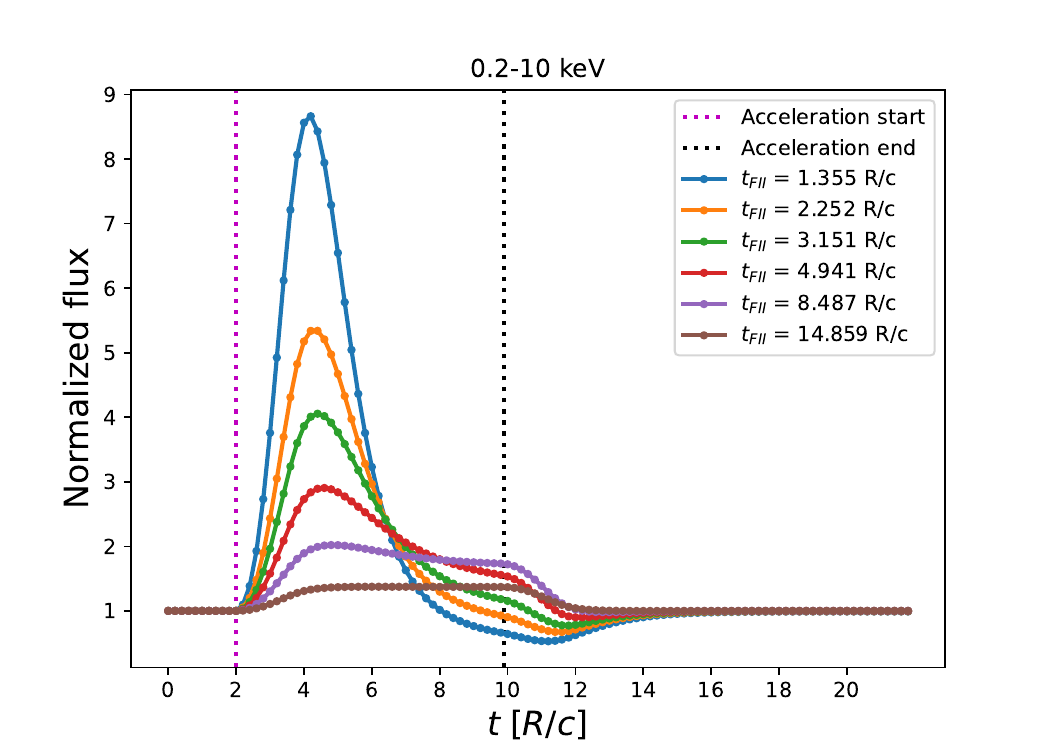}
    \caption{Optical and X-ray LC of flares resulting from hard-sphere Fermi II reacceleration for different timescales $t_{F_{II}}$ (scenario 3b).}
    \label{appfig:LCs_FII_reacc_compare}
\end{figure}

\FloatBarrier

\end{appendix}


\begin{thebibliography}{}

\bibitem[Abeysekara et al.(2020)]{Abeysekara2020} Abeysekara, A.~U., Benbow, W., Bird, R., et al.\ 2020, \apj, 890, 97 %doi:10.3847/1538-4357/ab6612

\bibitem[H.~E.~S.~S. Collaboration et al.(2019)]{Abdalla2019} H.~E.~S.~S. Collaboration, Abdalla, H., Adam, R., et al.\ 2019, \aap, 627, A159 %doi:10.1051/0004-6361/201935704


\bibitem[Abramowski et al. (2010)]{Abramowski2010} Abramowski,~A. et al. (H.E.S.S. Collaboration), 2010, A\&A, 520, A83

\bibitem[H.~E.~S.~S. Collaboration et al.(2012)]{Abramowski2012} H.~E.~S.~S. Collaboration, Abramowski, A., Acero, F., et al.\ 2012, \aap, 539, A149 %doi:10.1051/0004-6361/201117509


\bibitem[Acciari et al.(2020)]{Acciari2020} Acciari, V.~A., Ansoldi, S., Antonelli, L.~A., et al.\ 2020, \apjs, 248, 29 %doi:10.3847/1538-4365/ab89b5

\bibitem[Aimar et al. (2023)]{Aimar2023} Aimar N., Dmytriiev A., Vincent F.~H., et al.\, 2023, A\&A, 672, A62 %doi:10.1051/0004-6361/202244936


\bibitem[Albert et al. (2007)]{Albert2007} Albert,~J. et al. (MAGIC Collaboration), 2007, ApJ 669, 862

\bibitem[Asano \& Hayashida(2018)]{Asano2018} Asano, K. \& Hayashida, M.\ 2018, \apj, 861, 1, 31 %doi:10.3847/1538-4357/aac82a


\bibitem[Asano et al.(2014)]{Asano2014} Asano, K., Takahara, F., Kusunose, M., et al.\ 2014, \apj, 780, 64 %doi:10.1088/0004-637X/780/1/64

\bibitem[Baring et al.(2017)]{Baring2017} Baring, M.~G., B{\"o}ttcher, M., \& Summerlin, E.~J.\ 2017, \mnras, 464, 4875 %doi:10.1093/mnras/stw2344


\bibitem[Begelman et al. (1984)]{Begelman1984} Begelman, C. Mitchell et al., 1984, Rev. Mod. Phys. 56, 255-351

\bibitem[Bhatta \& Dhital(2020)]{Bhatta2020} Bhatta, G. \& Dhital, N.\ 2020, \apj, 891, 120 %doi:10.3847/1538-4357/ab7455

\bibitem[Biteau et al.(2020)]{Biteau2020} Biteau, J., Prandini, E., Costamante, L., et al.\ 2020, Nature Astronomy, 4, 124 %doi:10.1038/s41550-019-0988-4

\bibitem[B{\"o}ttcher(2019)]{Boettcher2019} B{\"o}ttcher, M.\ 2019, Galaxies, 7, 20 %doi:10.3390/galaxies7010020

\bibitem[Boettcher et al. (2019)]{Boettcher2019a} Böttcher,~M., Baring,~M.~G., 2019, ApJ, 887, 133

\bibitem[Boutelier et al. (2008)]{Boutelier2008} Boutelier,~T., Henri,~G., Petrucci,~P.~O., 2008, MNRAS, 390, L73

\bibitem[Britzen et al.(2018)]{Britzen2018} Britzen, S., Fendt, C., Witzel, G., et al.\ 2018, \mnras, 478, 3199 %doi:10.1093/mnras/sty1026


\bibitem[Casadio et al. (2015)]{Casadio2015} Casadio,~C. et al., 2015, ApJ, 813, 51

\bibitem[\protect\citeauthoryear{Cerruti}{2025}]{Cerruti2025} Cerruti M., 2025, A\&A, 698, A101 %doi:10.1051/0004-6361/202452289

\bibitem[Chang \& Cooper (1970)]{Chang1970} J. S. Chang, Cooper G., 1970, J. Comput. Phys. 6, Elsevier BV, 1-16

\bibitem[Chiaberge et al. (1999)]{Chiaberge1999}M. Chiaberge, G. Ghisellini, 1999, MNRAS 306, 551

\bibitem[Covino et al.(2019)]{Covino2019} Covino, S., Sandrinelli, A., \& Treves, A.\ 2019, \mnras, 482, 1270 %doi:10.1093/mnras/sty2720
   
\bibitem[Cox(1980)]{cox} Cox, J. P. 1980, Theory of Stellar Pulsation (Princeton University Press, Princeton) 165

\bibitem[Cox(1969)]{cox69} Cox, A. N.,\& Stewart, J. N. 1969, Academia Nauk, Scientific Information 15, 1

\bibitem[de Jaeger et al.(2023)]{deJaeger2023} de Jaeger, T., Shappee, B.~J., Kochanek, C.~S., et al.\ 2023, \mnras, 519, 6349 %doi:10.1093/mnras/stad060

\bibitem[Dmytriiev et al. (2021)]{Dmytriiev2021}A. Dmytriiev, H. Sol, A. Zech, 2021, MNRAS 505, 2712–2730

\bibitem[Dmytriiev \& B{\"o}ttcher(2024)]{Dmytriiev2024} Dmytriiev, A. \& B{\"o}ttcher, M.\ 2024, \aap, 687, A64 %doi:10.1051/0004-6361/202348269

\bibitem[Di Gesu et al.(2022)]{DiGesu2022} Di Gesu, L., Tavecchio, F., Donnarumma, I., et al.\ 2022, \aap, 662, A83 %doi:10.1051/0004-6361/202243168

\bibitem[Escudero Pedrosa et al.(2024)]{EscuderoPedrosa2024} Escudero Pedrosa, J., Agudo, I., Tramacere, A., et al.\ 2024, \aap, 682, A100 %doi:10.1051/0004-6361/202346885

\bibitem[\protect\citeauthoryear{Fossati et al.}{1998}]{Fossati1998} Fossati G., Maraschi L., Celotti A., Comastri A., Ghisellini G., 1998, MNRAS, 299, 433 %doi:10.1046/j.1365-8711.1998.01828.x

\bibitem[Geng et al.(2022)]{Geng2022} Geng, X., Ding, N., Cao, G., et al.\ 2022, \apjs, 260, 48 %doi:10.3847/1538-4365/ac64f6

\bibitem[Giannios et al. (2009)]{Giannios2009} Giannios,~D., Uzdensky,~D.~A., Begelman,~M.~C., 2009, MNRAS, 395, L29

\bibitem[Gould (1975)]{Gould1975}R. J. Gould, 1975, The Astrophysical Journal 196, 689-694

\bibitem[de Gouveia et al. (2010)]{Gouveia2010} de Gouveia Dal Pino,~E.~M., Piovezan,~P., Kadowaki,~L., Kowal,~G., Lazarian,~A., 2010, Highlights of Astronomy, 15, 247

\bibitem[Goyal et al.(2022)]{Goyal2022} Goyal, A., Soida, M., Stawarz, {\L}., et al.\ 2022, \apj, 927, 214 %doi:10.3847/1538-4357/ac4d95

\bibitem[Hovatta (2009)]{Hovatta2009}T. Hovatta, E. Valtaoja, M. Tornikoski, A. Lähteenmäki, Astron. Astrophys. 494, 527–537, 2009

\bibitem[Jormanainen et al.(2023)]{Jormanainen2023} Jormanainen, J., Hovatta, T., Christie, I.~M., et al.\ 2023, \aap, 678, A140 %doi:10.1051/0004-6361/202346286

\bibitem[Joshi et al.(2016)]{Joshi2016} Joshi, M., Marscher, A., \& B{\"o}ttcher, M.\ 2016, Galaxies, 4, 45 %doi:10.3390/galaxies4040045

\bibitem[Kakuwa et al.(2015)]{Kakuwa2015} Kakuwa, J., Toma, K., Asano, K., et al.\ 2015, \mnras, 449, 551 %doi:10.1093/mnras/stv281

\bibitem[Kapanadze et al.(2025)]{Kapanadze2025} Kapanadze, B., Gurchumelia, A., \& Aller, M.\ 2025, \apss, 370, 2, 17 %doi:10.1007/s10509-025-04411-0

\bibitem[Kardashev (1962)]{Kardashev1962}N. S. Kardashev, 1962, Soviet Astronomy 6

\bibitem[Katarzy{\'n}ski et al.(2006)]{Katarzynski2006} Katarzy{\'n}ski, K., Ghisellini, G., Mastichiadis, A., et al.\ 2006, \aap, 453, 47. doi:10.1051/0004-6361:20054176

\bibitem[Khatoon et al.(2024)]{Khatoon2024} Khatoon, R., B{\"o}ttcher, M., \& Prince, R.\ 2024, \apj, 974, 2, 233 %doi:10.3847/1538-4357/ad6f02

\bibitem[Kirk et al. (1998)]{Kirk1998} J.~G. Kirk, F. M. Rieger, A. Mastichiadis, Astron. Astrophys. 333, 452–458, 1998

\bibitem[Kowal et al. (2012)]{Kowal2012} G.~Kowal, E.~M.~de Gouveia Dal Pino, A.~Lazarian, 2012, Phys. Rev. Lett. 108, 241102

\bibitem[Kushwaha et al.(2014)]{Kushwaha2014} Kushwaha, P., Singh, K.~P., \& Sahayanathan, S.\ 2014, \apj, 796, 1, 61 %doi:10.1088/0004-637X/796/1/61

\bibitem[Kusunose et al.(2000)]{Kusunose2000} Kusunose, M., Takahara, F., \& Li, H.\ 2000, \apj, 536, 1, 299 %doi:10.1086/308928

\bibitem[Larionov et al. (2016)] {Larionov2016} Larionov,~V.~M., et al., 2016, MNRAS, 461, 3047

\bibitem[LeBihan et al. (2025)] {LeBihan2025} Le Bihan,~S., Dmytriiev,~A., Zech,~A., 2025, Proceedings of the 2024 Gamma Conference, astro-ph\/2501.14074

\bibitem[\protect\citeauthoryear{Lemoine et al.}{2019a}]{Lemoine2019a} Lemoine M., Gremillet L., Pelletier G., Vanthieghem A., 2019, PhRvL, 123, 035101 %doi:10.1103/PhysRevLett.123.035101

\bibitem[\protect\citeauthoryear{Lemoine et al.}{2019b}]{Lemoine2019b} Lemoine M., Vanthieghem A., Pelletier G., Gremillet L., 2019, PhRvE, 100, 033209 %doi:10.1103/PhysRevE.100.033209

\bibitem[\protect\citeauthoryear{Lemoine et al.}{2019c}]{Lemoine2019c} Lemoine M., Pelletier G., Vanthieghem A., Gremillet L., 2019, PhRvE, 100, 033210 %doi:10.1103/PhysRevE.100.033210

\bibitem[\protect\citeauthoryear{Lemoine \& Malkov}{2020}]{Lemoine2020} Lemoine M., Malkov M.~A., 2020, MNRAS, 499, 4972 %doi:10.1093/mnras/staa3131

\bibitem[Marscher(2014)]{Marscher2014} Marscher, A.~P.\ 2014, \apj, 780, 1, 87 %doi:10.1088/0004-637X/780/1/87

\bibitem[Melrose \& Pope(1993)]{Melrose1993} Melrose, D.~B. \& Pope, M.~H.\ 1993, \pasa, 10, 3, 222 %doi:10.1017/S1323358000025716

\bibitem[Lemoine et al.(2024)]{Lemoine2024} Lemoine, M., Murase, K., \& Rieger, F.\ 2024, \prd, 109, 063006 %doi:10.1103/PhysRevD.109.063006

\bibitem[Lewis et al.(2018)]{Lewis2018} Lewis, T.~R., Finke, J.~D., \& Becker, P.~A.\ 2018, \apj, 853, 6 %doi:10.3847/1538-4357/aaa19a

\bibitem[Liodakis (2015)]{Liodakis2015}I.Liodakis, V. Pavlidou, Mon. Not. R. Astron. Soc. 451, 3, 2434–2446, 2015

\bibitem[Luashvili et al.(2023)]{Luashvili2023} Luashvili, A., Boisson, C., Zech, A., et al.\ 2023, \mnras, 523, 404 %doi:10.1093/mnras/stad1393

\bibitem[Lucchini et al.(2019)]{Lucchini2019} Lucchini, M., Markoff, S., Crumley, P., et al.\ 2019, \mnras, 482, 4798 %doi:10.1093/mnras/sty2929


\bibitem[Marscher et al. (1985)]{Marscher1985} Marscher,~A.,~P., Gear,~W.~K., 1985, ApJ, 298, 114

\bibitem[Matthews et al.(2020)]{Matthews2020} Matthews, J.~H., Bell, A.~R., \& Blundell, K.~M.\ 2020, \nar, 89, 101543 %doi:10.1016/j.newar.2020.101543

\bibitem[Mizuno(1980)]{mizuno} Mizuno H. 1980, Prog. Theor. Phys., 64, 544

\bibitem[Moderski et al. (2005)]{Moderski2005}R. Moderski et al., 2005, Oxford University Press 363

\bibitem[Nalewajko et al.(2011)]{Nalewajko2011} Nalewajko,~K., et al., 2011, MNRAS 188

\bibitem[Nalewajko(2013)]{Nalewajko2013} Nalewajko, K.\ 2013, \mnras, 430, 2, 1324 %doi:10.1093/mnras/sts711

\bibitem[Otero-Santos et al.(2024)]{OteroSantos2024} Otero-Santos, J., Raiteri, C.~M., Acosta-Pulido, J.~A., et al.\ 2024, \aap, 686, A228 %doi:10.1051/0004-6361/202449647

\bibitem[\protect\citeauthoryear{Padovani \& Giommi}{1995}]{Padovani1995} Padovani P., Giommi P., 1995, ApJ, 444, 567 %doi:10.1086/175631

\bibitem[Paliya et al.(2015)]{Palyia2015} Paliya, V.~S., Sahayanathan, S., \& Stalin, C.~S.\ 2015, \apj, 803, 15 %doi:10.1088/0004-637X/803/1/15

\bibitem[Petropoulou et al.(2015)]{Petropoulou2015} Petropoulou, M., Piran, T., \& Mastichiadis, A.\ 2015, \mnras, 452, 3226 %doi:10.1093/mnras/stv1523

\bibitem[Potter \& Cotter(2012)]{Potter2012} Potter, W.~J. \& Cotter, G.\ 2012, \mnras, 423, 756 %doi:10.1111/j.1365-2966.2012.20918.x

\bibitem[Potter(2018)]{Potter2018} Potter, W.~J.\ 2018, \mnras, 473, 4107 %doi:10.1093/mnras/stx2371

\bibitem[Prince et al.(2018)]{Prince2018} Prince, R., Raman, G., Hahn, J., et al.\ 2018, \apj, 866, 16 %doi:10.3847/1538-4357/aadadb

\bibitem[Pushkarev et al. (2009)]{Pushkarev2009}A. B. Pushkarev, Y. Y. Kovalev, M. L. Lister, T. Savolainen, 2009, A\&A 507, L33

\bibitem[Raiteri et al. (2017)]{Raiteri2017} Raiteri,~C.~M. et al., 2017, Nature, 552, 374

\bibitem[Rajput et al.(2019)]{Rajput2019} Rajput, B., Stalin, C.~S., Sahayanathan, S., et al.\ 2019, \mnras, 486, 2, 1781 %doi:10.1093/mnras/stz941

\bibitem[\protect\citeauthoryear{Rieger}{2019}]{Rieger2019} Rieger F.~M., 2019, Galax, 7, 78 %doi:10.3390/galaxies7030078

\bibitem[\protect\citeauthoryear{Rieger \& Duffy}{2021}]{Rieger2021} Rieger F.~M., Duffy P., 2021, ApJL, 907, L2 %doi:10.3847/2041-8213/abd567

\bibitem[R{\"o}ken et al.(2018)]{Roeken2018} R{\"o}ken, C., Schuppan, F., Proksch, K., et al.\ 2018, \aap, 616, A172. doi:10.1051/0004-6361/201730622


\bibitem[Samarai et al.(2019)]{Samarai2019} Samarai, Imen Al and Alves Batista, R. and Barres de Almeida et al.\  2019, World Scientific %doi:10.1142/10986

\bibitem[Savolainen (2010)]{Savolainen2010}T. Savolainen, D. C. Homan, T. Hovatta, M. Kadler, Y. Y. Kovalev, M. L. Lister, E. Ros, J. A. Zensus, A\&A 512, A24, 2010

\bibitem[Sciaccaluga \& Tavecchio(2022)]{Sciaccaluga2022} Sciaccaluga, A. \& Tavecchio, F.\ 2022, \mnras, 517, 2502 %doi:10.1093/mnras/stac2755

\bibitem[Shukla \& Mannheim(2020)]{Shukla2020} Shukla, A. \& Mannheim, K.\ 2020, Nature Communications, 11, 4176 %doi:10.1038/s41467-020-17912-z

\bibitem[Sikora et al. (2001)]{Sikora2001} Sikora,~M., Blazejowski~M., Begelman,~M.~C., Moderski,~R., 2001, ApJ, 554, 1

\bibitem[Sol \& Zech(2022)]{SolZech2022} Sol, H. \& Zech, A.\ 2022, Galaxies, 10, 105 %doi:10.3390/galaxies10060105

\bibitem[Tammi et al. (2009)]{Tammi2009} Tammi~J., Duﬀy~P., 2009, MNRAS, 393, 1063

\bibitem[Tavecchio et al.(2022)]{Tavecchio2022} Tavecchio, F., Costa, A., \& Sciaccaluga, A.\ 2022, \mnras, 517, L16 %doi:10.1093/mnrasl/slac084

\bibitem[Terlevich(1992)]{terlevich} Terlevich, R. 1992, in ASP Conf. Ser. 31, Relationships between Active Galactic Nuclei and Starburst Galaxies, ed. A. V. Filippenko, 13

\bibitem[Tramacere et al.(2009)]{Tramacere2009} Tramacere, A., Giommi, P., Perri, M., et al.\ 2009, \aap, 501, 879 %doi:10.1051/0004-6361/200810865

\bibitem[Tramacere et al. (2011)]{Tramacere2011}A. Tramacere, E. Massaro, A. M. Taylor, 2011, Astrophys. J. 739, 66

\bibitem[Tramacere(2020)]{Tramacere2020} Tramacere, A.\ 2020, Astrophysics Source Code Library. ascl:2009.001

\bibitem[Tramacere et al. (2022)]{Tramacere2022}A. Tramacere, V. Sliusar, R. Walter, J. Jurysek, M. Balbo, 2022, A\&A 658, A173
   
\bibitem[Tscharnuter(1987)]{tscharnuter} Tscharnuter W. M. 1987, A\&A, 188, 55

\bibitem[\protect\citeauthoryear{Urry \& Padovani}{1995}]{Urry1995} Urry C.~M., Padovani P., 1995, PASP, 107, 803 %doi:10.1086/133630

\bibitem[Vieu et al.(2022)]{Vieu2022} Vieu, T., Gabici, S., \& Tatischeff, V.\ 2022, \mnras, 510, 2, 2529 %doi:10.1093/mnras/stab3564

\bibitem[Yorke(1980a)]{yorke80a} Yorke, H. W. 1980a, A\&A, 86, 286
      
\bibitem[Yu et al.(2024)]{Yu2024} Yu, J., Ding, N., Fan, J., et al.\ 2024, \apj, 967, 2, 96 %doi:10.3847/1538-4357/ad3e68

\bibitem[Zech \& Lemoine(2021)]{Zech2021} Zech, A. \& Lemoine, M.\ 2021, \aap, 654, A96 %doi:10.1051/0004-6361/202141062

\bibitem[Zhang et al.(2016)]{Zhang2016} Zhang, H., Diltz, C., \& B{\"o}ttcher, M.\ 2016, \apj, 829, 2, 69 %doi:10.3847/0004-637X/829/2/69

\bibitem[Zheng(1997)]{zheng} Zheng, W., Davidsen, A. F., Tytler, D. \& Kriss, G. A. 1997
\end{thebibliography}
\end{document}